\documentclass[pra,letterpaper,aps,10pt,superscriptaddress,twocolumn,floatfix,showpacs]{revtex4-1}
\usepackage{amsmath,amsfonts,dsfont,graphicx,amssymb,microtype,braket,xcolor,upgreek}
\DeclareGraphicsExtensions{.pdf}
\usepackage[colorlinks, linkcolor=blue, citecolor=blue, urlcolor=blue, breaklinks=true]{hyperref}
\newcommand{\ts}[1]{_\text{#1}}
\newcommand{\me}{\mathrm{e}}
\newcommand{\mi}{\mathrm{i}}
\newcommand{\dd}{\mathrm{d}}

\newcommand{\bk}{\mathbf{k}}
\newcommand{\bq}{\mathbf{q}}

\newcommand{\br}{\mathbf{r}}
\newcommand{\bb}{\boldsymbol{\beta}}
\newcommand{\be}{\boldsymbol{\eta}}
\newcommand{\bO}{\boldsymbol{\Omega}}
\newcommand{\bG}{\boldsymbol{\Gamma}}
\newcommand{\bM}{\boldsymbol{\mathcal{M}}}
\newcommand{\calA}{\mathcal{A}}
\newcommand{\bI}{\boldsymbol{\mathcal{I}}}
\newcommand{\bS}{\boldsymbol{\mathcal{S}}}
\newcommand{\bT}{\boldsymbol{\mathcal{T}}}

\usepackage{placeins}
\newcommand{\iu}{\mathrm{i}}
\newcommand{\diff}{\mathrm{d}}
\usepackage{mathtools}
\usepackage{physics}

\newcommand{\WN}{for $\mathcal{N}=40^2=1600$ emitters}
\newcommand{\wfWN}{$\mathcal{N}=40^2=1600$ emitters}

\newcommand{\MPL}{Max  Planck  Institute  for  the  Science  of  Light,  Staudtstra{\ss}e  2,  D-91058  Erlangen,  Germany}
\newcommand{\FAU}{Department of Physics, Friedrich-Alexander-Universit\"{a}t Erlangen-N\"urnberg, Staudtstra{\ss}e 7, D-91058 Erlangen, Germany}

\binoppenalty=\maxdimen
\relpenalty=\maxdimen

\begin{document}

\title{Linear optical elements based on cooperative subwavelength emitter arrays}
\author{Nico S. Ba\ss ler}
\affiliation{\FAU}
\affiliation{\MPL}
\author{Michael Reitz}
\affiliation{\MPL}
\author{Kai Phillip Schmidt}
\affiliation{\FAU}
\author{Claudiu Genes}
\affiliation{\MPL}
\affiliation{\FAU}
\date{\today}

\begin{abstract}
  We describe applications of two-dimensional subwavelength quantum emitter arrays as efficient optical elements in the linear regime. For normally incident light, the cooperative optical response, stemming from emitter-emitter dipole exchanges, allows the control of the array's transmission, its resonance frequency, and bandwidth. Operations on fully polarized incident light, such as generic linear and circular polarizers as well as phase retarders can be engineered and described in terms of Jones matrices. Our analytical approach and accompanying numerical simulations identify optimal regimes for such operations and reveal the importance of adjusting the array geometry and of the careful tuning of the external magnetic fields amplitude and direction.
\end{abstract}


\maketitle
\section{Introduction}
\begin{figure*}[t]
  \centering
  \includegraphics[width=0.9\textwidth]{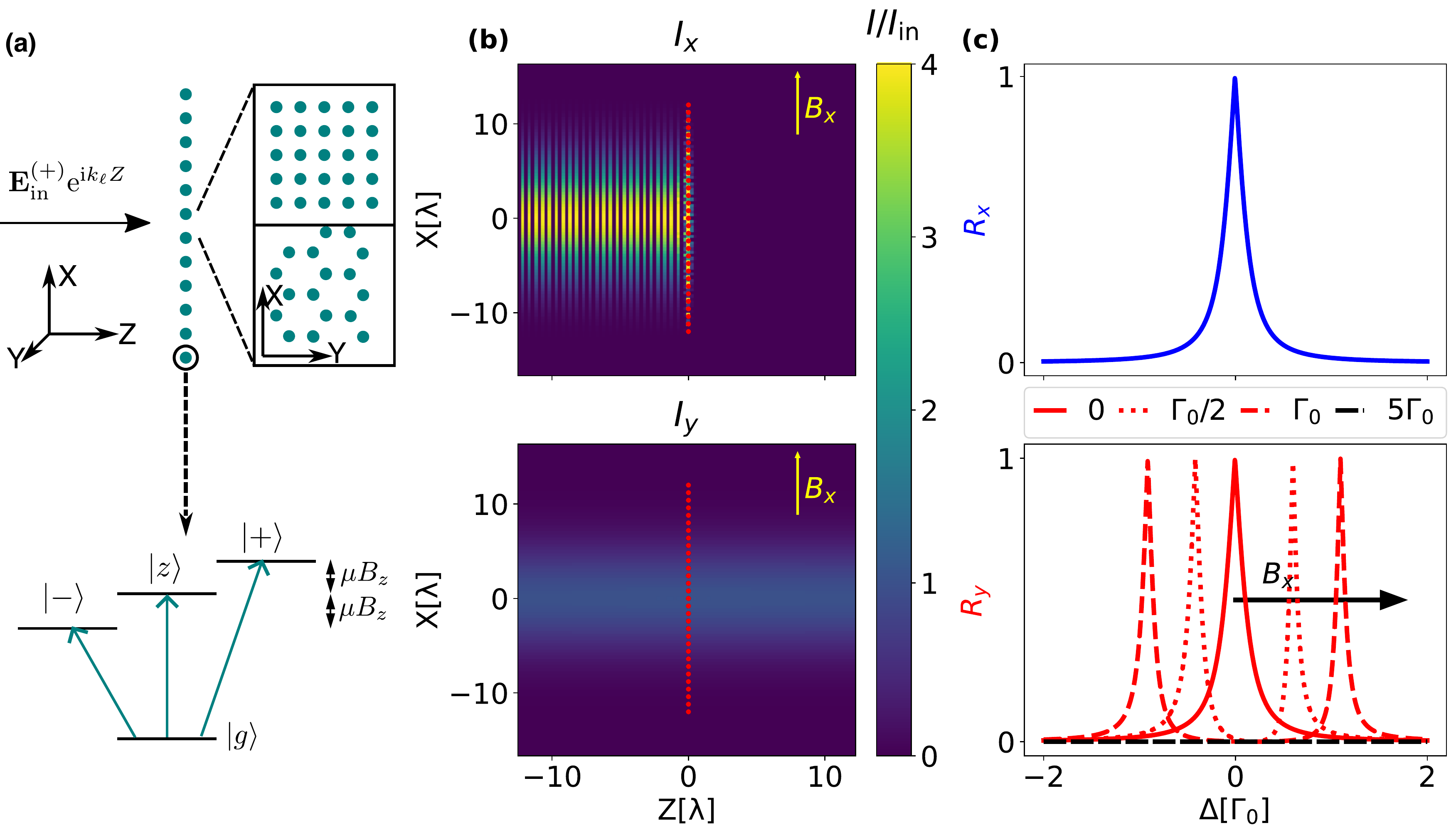}%
  \caption{(a) An incoming plane wave propagating along the $z$ direction is impinging, at normal incidence, on a two-dimensional subwavelength quantum emitter array. The inset shows two possible arrangements of the emitters comprising the metasurface. The lower panel shows the internal electronic structure of each emitter, where the excited electronic states can be split in energy by an externally controlled magnetic field (here pointing in the $z$ direction). (b) Field intensity distribution for the $x$ polarization (upper plot) and $y$ polarization (lower plot) for a metasurface acted upon with a magnetic field in the $x$ direction. The array acts as a perfect polarizer, fully reflecting the $x$-polarized light while allowing full transmission for the orthogonal $y$-polarization component. (c) Scan of the reflectivity for the $x$-polarized light (upper plot) and $y$-polarized light (lower plot) as a function of the incoming field frequency (quantified by the detuning $\Delta$ with respect to the bare array resonance). An increasing magnetic field in the $x$ direction allows for the progressive shift of the reflection peak for the $y$ polarization far enough from the reflectivity window, thus allowing for the implementation of a perfect polarizer.}
  \label{fig1}
\end{figure*}

Subwavelength arrays of emitters react cooperatively to coherent laser illumination, owing to the strong intrinsic near-field dipole-dipole interactions~\cite{chang2018quantum, asenjogarcia2017exponential, reitz2022cooperative, jenkins2012controlled}. In essence, their linear optical response is similar to that of arrays of classical dipoles, such as is the case for plasmonic lattices~\cite{solntsev2021metasurfaces, yao2014plasmonic, kravets2018plasmonic, binalam2021ultra, jenkins2017manybody}, in the fact that light couples to delocalized surface excitations with tunable resonances. In addition, the linewidths of these collective resonances can also be tuned: the collective coupling of all emitters to the ubiquitous electromagnetic vacuum, in which they are inherently embedded, leads to super- or subradiant behavior, i.e.,~spontaneous emission rates smaller or larger than the independent emitter radiative rate \cite{dicke1954coherence, lehmberg1970radiation}. Theoretical and experimental proposals are currently exploring the fascinating properties of emitter arrays with potential use as metasurfaces which can e.g.~exhibit perfect reflectivity~\cite{bettles2016enhanced, Shahmoon2017,rui2020asubradiant, ballantine2020optical, ballantine2021cooperative}, for applications as photon storage devices and in quantum information processing~\cite{plankensteiner2015selective, facchinett12016storing, manzoni2018optimization, guimond2019subradiant, Bekenstein2020}, or as atom-thick membranes with extremely small mass, useful for optomechanical applications \cite{shahmoon2019collective, shahmoon2020quantum}. In the direction of topological quantum optics~\cite{bettles2017topological,Perczel2017,Perczel2017_2,Perczel2020}, it has been proposed to combine the subradiant protection of excitation against radiative losses with the topological protection of excitation against structural defects. The intrinsic nonlinearity of electronic transitions adds further possibilities into applications in nonlinear optics~\cite{morenocardoner2021quantum, rusconi2021exploiting, srakaew2022subwavelength}. Possible experimental implementations of quantum emitter arrays include atoms trapped in optical lattices \cite{bloch2005ultracold, rui2020asubradiant} as well as localized excitons in semiconductor structures \cite{Palacios2017largescale, Li2021scalable} where each platform is subject to different kinds of defects and disorder.\\
\indent Here, we theoretically characterize the functionality of two-dimensional subwavelength emitter arrays as linear optical elements, such as polarizers or waveplates. More specifically, we show that tuning of the array geometry combined with the application of an externally controllable magnetic field allows for the implementation of quasi general Jones matrices. For fully polarized incident light, such $2\times 2$ matrices describe polarizers, i.e.,~filters for light polarization perpendicular to some chosen axis, or phase retarders exhibiting controllable delays $\me^{\iu\phi_x}$ and $\me^{\iu\phi_y}$ for the orthogonal $x$- and $y$-polarization components. The main tuning knob revealed by our theoretical analysis is the possibility of manipulating the band structure of the collective excitations, residing on the surface of the array, by the control of the magnitude and direction of an externally applied magnetic field. In particular, for the simplified situation of normal, uniform plane-wave illumination, only the zero-momentum mode surface excitation is driven, i.e.,~a symmetric collective mode of all emitter dipoles oscillating in phase. In such a case, the tuning up of the magnetic field allows to create a large gap between the optimal points of the reflection windows for orthogonal polarizations. For example, a large magnetic field in the $x$ direction can help shift the reflectivity window of the $y$-polarized light thus allowing it to be completely transmitted while at the same time fully reflecting the $x$-polarization component. The situation is roughly depicted in Fig.~\ref{fig1} which illustrates the array geometry, the internal electronic structure of an individual quantum emitter, the field intensity for the two orthogonal polarizations and the frequency gap in transmission peaks between the two polarizations, introduced by a controllable externally applied magnetic field.\\
%%%%%%%%%%%%%%%%%%%%%
%%%%%%%%%%%%%%%%%%
\indent The analytical procedure involves the following steps: i) connecting the outgoing, far field to the cooperative response of the emitters, ii) computing the response of the array to the incident field in both real space and Fourier domains and iii) connecting the total outgoing, transmitted field to the incoming field via a transmission matrix such that $\mathbf E^{(+)}(Z>0)=\bT \mathbf E^{(+)}_\text{in}(Z<0)$. On the way, a reduction from a three-component treatment to two components is performed owing to the subwavelength nature of the array, which imposes that the $z$-polarized component of the outgoing field always vanishes. This allows the derivation of an analytical expression of a $2\times 2$ Jones matrix directly connecting the two components of the outgoing field to the ones of the incoming field \cite{hecht2017optics, chekhova2021polarization}. These analytical expressions allows to identify the tuning knobs for the on-demand implementation of a subset of polarizers or waveplates and to manipulate the effective macroscopic polarizability tensor of the array.\\
\indent The paper is organized as follows: The analytical formalism is introduced in Sec.~\ref{Sec2}. The far field emitted as a cooperative response to normal incident illumination is connected to the dipolar response of the array in the weak driving limit. The array response is computed with an open system approach, in the form of a master equation, to light-matter interactions. In Sec.~\ref{Sec3}, the reduction to a two-by-two description is detailed which allows then for the derivation of a general expression for the Jones matrices under the action of an external magnetic field. The implementation procedure for polarizers and waveplates is detailed in Secs.~\ref{Sec3B} and~\ref{Sec3C}. Additional aspects relevant for experimental implementations such as the effect of thermal motion, defects and the departure from the weak excitation condition are discussed in Sec.~\ref{Sec4}.\\
%%%%%%%%%%%%%%%%%%%%%%%%%%%%%%%%%%%%%%%%%%%%%%%%%%%%%%%%%%
%%%%%%%%%%%%%%%%%%%%%%%%%%%%%%%%%%%%%%%%%%%%%%%%%%%%%%%%%%
\section{Linear optical response of a two-dimensional array}
\label{Sec2}
%%%%%%%%%%%%%%%%%%%%%%%%%%%%%%%%%%%%%%%%%%%%%%%%%%%%%%%%%%
%%%%%%%%%%%%%%%%%%%%%%%%%%%%%%%%%%%%%%%%%%%%%%%%%%%%%%%%%%
We consider the situation depicted in Fig.~\ref{fig1}(a) showing a fully polarized incident field with wave vector $k_\ell \hat{e}_z$, frequency $\omega_\ell=2\pi c/\lambda$ (with wavelength $\lambda=2\pi/k_\ell$) and electric field amplitudes $E_{\text{in},x}$ and $E_{\text{in},y}$ (such that the total polarization points in some direction $\theta$ in the $xy$ plane). The field impinges, at normal incidence, upon a subwavelength quantum emitter array located in the $xy$ plane at $z=0$. Later, we will generalize the approach to oblique incidence. The array is comprised of $\mathcal{N}$ identical emitters, each with an internal structure described by a $J=0$ to $J=1$ transition, such that an externally applied magnetic field $\mathbf{B}$ can split the degeneracy of the three excited sublevels by Zeeman frequency shifting. This is depicted in the lower panel of Fig.~\ref{fig1}(a). Denoting the four states by $\ket{g}$ and $\ket{\nu}$ (with $\nu=0,\pm$), the transition dipole operator for each emitter can be written as $\mathbf{d}=\sum_\nu \mathbf{d}_{\nu}\sigma_\nu+\mathrm{h.c.}$ where $\mathbf{d}_{\nu}=\bra{g}\mathbf{d}\ket{\nu}$ and $\sigma_\nu=\ket{g}\bra{\nu}$ is the corresponding lowering operator for the transition. Alternatively, we can define combinations $\sigma_x=\sigma_{-}+\sigma_{+}$ and $\sigma_y=\iu(\sigma_{-}-\sigma_{+})$ and rewrite $\mathbf{d}=\mathbf{d}^{(+)}+\mathbf{d}^{(-)}$, in terms of positive and negative frequency components. The positive frequency component is then expressed in a Cartesian basis as $\mathbf{d}^{(+)}=d_x\sigma_x \hat{e}_x+d_y\sigma_y\hat{e}_y+d_z\sigma_z\hat{e}_z$ (with the negative component obtained as its Hermitian conjugate). We will denote by the index $\alpha=x,y,z$ any components in the Cartesian basis.\\

%%%%%%%%%%%%%%%%%%%%%%%%%%%%%%%%%%
\subsection{Dynamics of a driven array}
%%%%%%%%%%%%%%%%%%%%%%%%%%%%%%%%%%
We assume that the frequency splitting (for the degenerate case, in the absence of an externally applied magnetic field) for all 3 possible optical transitions is $\omega_0$ ($\hbar=1$) which is detuned by $\Delta=\omega_\ell-\omega_0$ from the laser frequency. Denoting now each lowering operator in the Cartesian basis as $\sigma_{j,\alpha}$ (for the $\alpha$ transition within the particular emitter $j$) we can write the Hamiltonian of a driven emitter in the frame of the laser as a sum between the free evolution part $\mathcal{H}_0$ and the driving part $\mathcal{H}_\ell$
\begin{align}
\mathcal{H}_0+\mathcal{H}_\ell=-\sum_{j,\alpha}\Delta \sigma_{j,\alpha}^\dagger\sigma_{j,\alpha}^{\phantom{\dagger}}+\sum_{j,\alpha}\left(\eta_\alpha \sigma_{j,\alpha}^\dagger+\eta_{\alpha}^* \sigma_{j,\alpha}^{\phantom{\dagger}}\right).
\end{align}
We have assumed constant, uniform illumination such that the Rabi frequencies $\eta_x=d_x E_{\text{in},x}$, $\eta_y=d_y E_{\text{in},y}$, and $\eta_z=0$ are independent of the emitter position.\\
As the emitters are closely positioned with respect to each other, dipole-dipole shifts will occur, which can be encompassed in the following Hamiltonian
\begin{align}
\mathcal{H}_\text{d-d}=\sum_{j, j', \alpha,\alpha'}\Omega^{\alpha,\alpha'}_{jj'}\sigma_{j,\alpha}^\dagger\sigma_{j',\alpha'}^{\phantom{\dagger}},
\end{align}
describing an exchange of excitation between two emitters indexed by $j$ and $j'$ and between transitions $\alpha$ and $\alpha'$. This can be derived from the general free space photonic Green's tensor (see Appendix \ref{AppendixA} for more details)
\begin{align}
\label{definitiongreen}
\mathbf{G}(\mathbf{R})=\left(\mathds{1}+\frac{1}{k_0^2} \nabla\otimes\nabla\right)\frac{\me^{\mi k_0R}}{4\pi R}-\frac{\mathds{1}}{3k_0^2}\delta(\mathbf{R}),
\end{align}
where $\omega_0=ck_0$, the symbol $\otimes$ denotes the dyadic product, $R=|\mathbf{R}|$ and the last term removes the divergence on the self interaction terms at $\mathbf{R}=0$. The exchange rate $\Omega^{\alpha,\alpha'}_{jj'}$ is then given as \cite{gruner1996green,dung2002resonant, buhmann2007dispersion}
\begin{align}
\Omega^{\alpha,\alpha'}_{jj'}\equiv\Omega^{\alpha,\alpha'}(\br_{jj'})=-\mu_0 \omega_0^2 \,\mathbf{d}_{\alpha}^*\cdot\text{Re}[ \mathbf{G}(\mathbf{r}_{jj'})]\cdot\mathbf{d}_{\alpha'}^{\phantom{*}},
\end{align}
which strongly depends on the inter-emitter separation $\mathbf{r}_{jj'}=\mathbf{r}_{j}-\mathbf{r}_{j'}$ and scales as $|\br_{jj'}|^{-3}$ in the near-field region.\\
\indent In addition, an externally applied magnetic field $\mathbf{B}$ adds to the Hamiltonian by differentially shifting the excited levels with magnetic moment $\mu$. For example, the $z$ component $B_z$ shifts the $\ket{\pm}$ levels by $\pm\Delta_B^{(z)}=\pm \mu B_z$. In the Cartesian basis, this can be reexpressed as
\begin{align}
\mathcal{H}_\text{B}^{(z)}=\mi \sum_{j} \Delta_B^{(z)} (\sigma_{j,y}^\dagger \sigma_{j,x}^{\phantom{\dagger}}-\sigma_{j,x}^\dagger \sigma_{j,y}^{\phantom{\dagger}}),
\end{align}
which can be seen as a mixing or excitation hopping between different levels within the same emitter. The expression for $\mathcal{H}_\text{B}^{(x,y)}$ is similar with the cyclic permutation of indexes (see Appendix \ref{AppendixB}).\\
\indent To the coherent processes in the Hamiltonian, one adds the collective decay imposed by the interaction with the electromagnetic vacuum
\begin{align}
\mathcal{L}[\rho]= \sum_{j,j',\alpha,\alpha'}\Gamma^{\alpha, \alpha'}_{jj'}\left[\sigma_{j,\alpha}^{\phantom{\dagger}}\rho\sigma_{j',\alpha'}^\dagger-\frac{1}{2}\left\{\sigma_{j,\alpha}^\dagger\sigma_{j',\alpha'}^{\phantom{\dagger}},\rho\right\}\right],
\end{align}
which describes a superoperator acting on the density operator of the system $\rho$.
The decay terms are also obtained from the Green's tensor by \cite{gruner1996green,dung2002resonant, buhmann2007dispersion}
\begin{align}
\Gamma^{\alpha, \alpha'}_{jj'}\equiv\Gamma^{\alpha,\alpha'}(\br_{jj'})=2\mu_0 \omega_0^2\,\mathbf{d}_{\alpha}^*\cdot\text{Im}[ \mathbf{G}(\mathbf{r}_{jj'})]\cdot\mathbf{d}_{\alpha'}^{\phantom{*}}.
\end{align}
The independent decay of an individual emitter is given by $\Gamma_{jj}^{\alpha,\alpha'}=\Gamma_0\delta_{\alpha,\alpha'}$ with the spontaneous emission rate $\Gamma_0=\omega_0^3 d^2/(3\pi\epsilon_0 c^3)$ (we assume all transitions to have identical dipole moment  $d_\alpha=d$ in the following). The dynamics of the system is then described by an open system master equation for $\rho$ which reads
\begin{align}
\frac{\dd \rho}{\dd t}=\mi [\rho, \mathcal{H}_0+\mathcal{H}_\ell+\mathcal{H}_\text{d-d}+\mathcal{H}_\text{B}]+\mathcal{L}[\rho].
\end{align}
We will make use of this master equation in Sec.~\ref{Sec2C} to compute the linear optical response as quantified by a macroscopic polarizability tensor for the whole two-dimensional array.

%%%%%%%%%%%%%%%%%%%%%%%%%%%%%%%%%%
\subsection{The far radiated field}
%%%%%%%%%%%%%%%%%%%%%%%%%%%%%%%%%%
The field emitted by a two-dimensional metasurface comprised of $\mathcal{N}$ emitters is the sum of the individually emitted dipole fields. However, the high density, i.e.,~small emitter-emitter separations, resulting in collective phase shifts and collective dissipation, will give rise to a more complex, cooperative response. Let us denote the individual emitters by an index $j$ and compute the classical electric field amplitude in the far field at some position on the $z$ axis at distance $Z>0$. For simplicity, we restrict the discussion to uniform illumination at normal incidence, with a plane wave of positive frequency component $\mathbf{E}_{\text{in}}^{(+)} \me^{\mi k_\ell z}$. The input field can be written as a three-component vector $(E_{\text{in},x},E_{\text{in},y},0)^\top$. The far field vector is then related to the array response as
\begin{equation}
\label{eq::farfield}
  \mathbf{E}^{(+)}(Z) = \mathbf{ E}_{\text{in}}^{(+)} \me^{\mi k_\ell Z}+\frac{3\pi\Gamma_0}{k_0 d}\sum_{j=1}^{\mathcal{N}} \mathbf{G}^\text{far}(Z \hat{e}_z-\mathbf{r}_j) \boldsymbol{\beta}_j,
\end{equation}
where each of the $\boldsymbol{\beta}_j=\langle \boldsymbol{\sigma}_j\rangle$ is a three-component vector, containing the Cartesian components of the expectation values of the lowering operators $\boldsymbol{\sigma}_j$. The sum above applies to all the emitters in the array situated at positions $\mathbf{r}_j$. The dipolar response is computed via the far-field component of the Green's tensor $\mathbf{G}^\text{far}(Z \hat{e}_z-\mathbf{r}_j)$, which is a simplified version of the already invoked Green's tensor under the assumption that $k_0 Z\gg 1$. Generally, for an arbitrary array, the dipole-emitted field could have a component in the direction of propagation. As it will be shown in Sec.~\ref{Sec3}, in the particular case of subwavelength arrays, the far field contains only $xy$ components, even if the array dipoles have a $z$ component.\\
%%%%%%%%%%%%%%%%%%%%%%%%%%%%%%%%%%%%%%
\subsection{Surface modes of the array}
\label{Sec2C}
%%%%%%%%%%%%%%%%%%%%%%%%%%%%%%%%%%%%%%

In the low-excitation limit where each electronic transition is only weakly excited $\langle{\sigma_{j,\alpha}^z\rangle}\approx -1$, the response of the emitter array can be computed from the master equation approach to open quantum systems, as detailed in the previous subsection. The coupling of the incident field to the electronic transitions is quantified by the vector of Rabi frequencies $\boldsymbol{\eta}$ with components ($\eta_x,\eta_y,0)$. We will proceed in analyzing the dipolar response both in the position space and via a Fourier transform allowing the understanding in terms of quasi-momenta residing in the reciprocal space. In the position space, one can derive a set of equations of motion for the three-component coherence of each $j$-indexed emitter
\begin{align}
\dot{\boldsymbol{\beta}}_j=-\mi\sum_{j'}\mathbf{M}_{jj'}\boldsymbol{\beta}_{j'}-\iu\boldsymbol{\eta}.
\end{align}
The matrix $\mathbf{M}_{jj'}=-\Delta\mathds{1}\delta_{j,j'}+\bO_{jj'}-\mi\bG_{jj'}/2+\mathbf{M}_\text{B}\delta_{j,j'}$ contains the laser detuning,  the effect of the coherent and incoherent interactions induced by the electromagnetic vacuum and the couplings introduced by the magnetic field in the Cartesian basis (explicitly specified in Appendix \ref{AppendixD}). Under the assumption of uniform illumination, all emitters are identically excited in steady-state $\boldsymbol{\beta}_j=\boldsymbol{\beta}$ with the vector of amplitudes
\begin{equation}
\label{eq:beta}
\boldsymbol{\beta}=-\bM^{-1}\boldsymbol{\eta},
\end{equation}
where the matrix $\bM=\sum_{j'}\mathbf{M}_{0j'}$ does not depend on the index $j$ for an infinite array which is why we set it to zero. This expression will later allow us to simply compute the response of the array in the next section by replacing it into Eq.~\eqref{eq::farfield}.  Notice that $\bM$ is a $3\times 3$ matrix and even if the incoming field has a vanishing $z$ component, the dipole pattern imprinted onto the array can have a $z$ component.\\
%%%%%%%%%%%%%%%%%%%%%%%%%%%%%%%%%%
\begin{figure}
  \centering
  \includegraphics[width=\columnwidth]{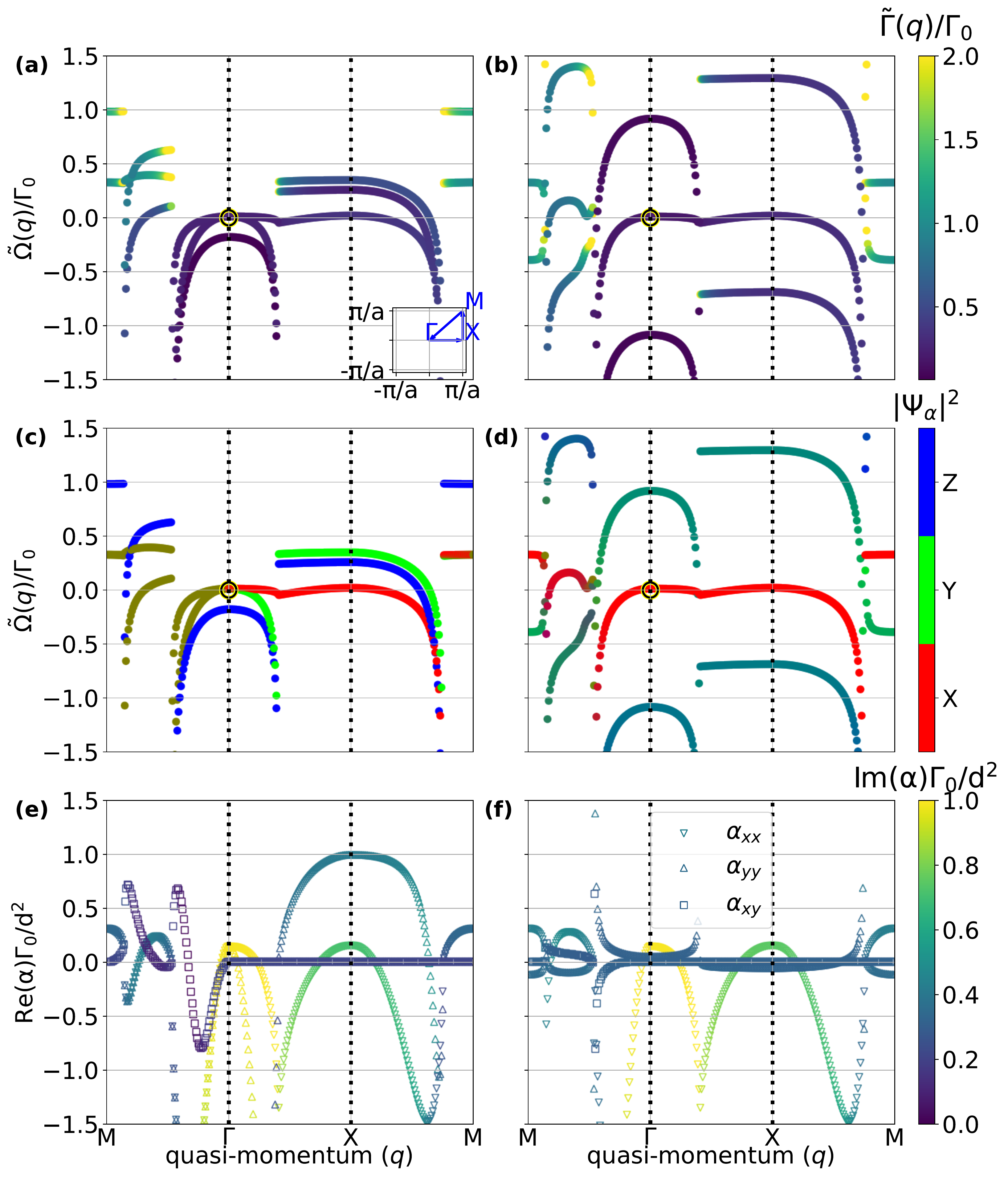}
  \caption{(a)-(d) Photonic band structures (eigenvalues of $\mathbf{\boldsymbol{\mathcal{M}}}(\bq)$ for $\Delta=0$) and (e)-(f) polarizability of the emitter array in the absence (left column) and with an applied magnetic field in $x$ direction (right column) for a square lattice with spacing $a=0.8\lambda$. The inset in (a) shows the path taken between points in the 2D reciprocal space. (a) Real part of the photonic band structure with color coding indicating the corresponding imaginary part, i.e.,~the decay rate which determines how sub- or superradiant a mode is at the particular quasi-momentum. For $a=0.8\lambda$, the decay rate at the $\Gamma$ point is subradiant for instance. The pointer indicates the perfect reflectivity point situated at the $\Gamma$ point for both incoming field polarizations. In (b), the magnetic field shifts the reflectivity point for the $y$ component but not for the $x$ component. This is more evident in (c) and (d) where the color coding indicates the overlap of the eigenstate with the Cartesian components such that the RGB color is defined as $(\abs{\Psi_x}^2,\abs{\Psi_y}^2,\abs{\Psi_z}^2)$. The applied magnetic field hybridizes the $yz$ bands  and shifts the resonances of these bands away from the zero energy point. In (e) and (f) we illustrate the real part and, in the color scale, the imaginary part of certain components $\alpha_{xx}$,  $\alpha_{yy}$, and $\alpha_{xy}$ of the polarizability tensor. The applied magnetic field for (b), (d), and (f) is $\mu B_x=\Gamma_0$.}
  \label{fig:surface_modes}
\end{figure}
%%%%%%%%%%%%%%%%%%%%%%%%
\indent The solution discussed in the previous paragraph corresponds to uniform, perpendicular illumination, where only the symmetric surface mode is activated. More generally, for non-normal incidence of the laser with a wave vector $ \bk_\parallel$ parallel to the array plane, the equations of motion can be expressed in Fourier space as
\begin{align}
\label{eq:betaq}
\dot{\tilde{\bb}}_\bq=-\mi\mathbf{\boldsymbol{\mathcal{M}}}(\bq)\tilde{\bb}_\bq-\mi\mathcal{N} \be\delta_{\bq, \bk_\parallel},
\end{align}
where the evolution matrix is now given by $\mathbf{\boldsymbol{\mathcal{M}}}(\bq)=-\Delta\mathds{1}+\tilde{\bO}(\bq)-\mi\tilde{\bG}(\bq)/2+\mathbf{M}_\text{B}$ and contains the matrices describing the effective frequency shifts and decay rates as a function of the in-plane momentum  $\tilde{\bO}(\bq)-\mi\tilde{\bG}(\bq)/2$ which are modified by the presence of the magnetic field $\mathbf{M}_\text{B}$. Diagonalization of the matrix $\boldsymbol{\mathcal{M}}(\bq)$ gives rise to the photonic band structure of the array (see paragraph below). Eq.~\eqref{eq:betaq} indicates that only the mode corresponding to $\bq=\bk_\parallel$ is driven by the laser. The symmetric illumination scenario discussed in the previous paragraph then simply refers to $\bq=0$.
The real and imaginary parts of the matrices describing the effective frequency shift and decay rate of a given $\mathbf q$ mode of the array are obtained by the Fourier transforms $\tilde{\bO}(\mathbf q)=\sum_{\br \in \Lambda} \me^{-\iu \mathbf q \cdot \br}\bO (\br)$ and $\tilde{\bG}(\mathbf q)=\sum_{\br \in \Lambda} \me^{-\iu \mathbf q \cdot \br}\bG (\br)$
involving a summation over all vectors of the lattice $\mathbf{r}\in \Lambda$. This can be compactly expressed in terms of the Green's tensor as
\begin{align}
\tilde{\bO}(\mathbf q)-\iu \frac{\tilde{\bG}(\mathbf q)}{2}=-\frac{3}{2}\Gamma_0\lambda_0\sum_{\br \in \Lambda}  \me^{-\iu \mathbf q\cdot \br} \mathbf{G}(\mathbf{r}).
\end{align}
\indent While analytical expressions for the dipole-dipole interactions are not available analytically, for subwavelength arrays excited at normal incidence, the relevant decay rate matrix at zero quasi-momentum can be approximated by (for derivation see Appendix \ref{AppendixC})
\begin{align}
\tilde{\mathbf{\Gamma}}(0)=\frac{3\Gamma_0}{4\pi}\frac{\lambda_0^2}{\calA}\begin{pmatrix}
1 & 0 & 0 \\
0 & 1 & 0 \\
0 & 0 & 0
\end{pmatrix},
\end{align}
where $\calA$ is the area of the unit cell of the real-space lattice. This result implies that for the symmetric mode of subwavelength lattices, spontaneous emission of the dipoles along the $z$ direction is completely suppressed while it assumes the simple expression $\tilde{\Gamma}(0)=3(\Gamma_0/{4\pi})(\lambda_0^2/\calA)$ for the other two directions.\\
\indent The modes of the array define the response to external illumination via the polarizability tensor. This can be expressed in the Fourier domain simply as $\boldsymbol\alpha(\bq)=d^2\mathbf{\boldsymbol{\mathcal{M}}}^{-1}(\bq)$. For a single incident polarization this assumes a simple form derived in Refs.~\cite{bettles2016enhanced, Shahmoon2017} while here, owing to its cross-coupling components, it can describe changes in one polarization owed to the other orthogonal one. The photonic band structure and polarizability are illustrated  in Fig.~\ref{fig:surface_modes} in the absence (left column) and with an applied magnetic field (right column). The photonic band structure is obtained from the diagonalization of the matrix $\boldsymbol{\mathcal{M}}(\bq)$, giving rise to three energy bands (for lattices containing a single emitter per unit cell). While in Figs.~\ref{fig:surface_modes}(a),(b) the color coding indicates the decay rate, in Figs.~\ref{fig:surface_modes}(c),(d) the color shows the relative $x/y/z$ content of a respective mode. In the absence of an external field, at the $\Gamma$ point (corresponding to zero quasi-momenta both in the $x$ and $y$ directions) the $x$  and $y$ bands are completely degenerate, owing to the symmetry of the square lattice, which is also reflected in the fact that the cross term of the polarizability tensor $\alpha_{xy}$ is zero, implying that the bands do not hybridize. The $z$ band does not have this symmetry constraint and its resonance is at a different frequency at the $\Gamma$ point. Going away from the $\Gamma$ point, the $xy$ bands start to hybridize slightly leading to different band shapes. Different bands for $x$ and $y$ also emerge due to the breaking of the $xy$ symmetry by picking a wave vector which does not obey this symmetry. Turning on a magnetic field in the $x$ direction has dramatic effects on all bands except the $x$ band which is pure as seen in Fig.~\ref{fig:surface_modes}(b),(d). Indeed, the strong hybridization of the $y$  and $z$ band induced by the external magnetic field in $x$ direction leads to a strong splitting of these bands around $\tilde\Omega(\mathbf q)=0$ so that a laser with frequency $\omega_0$ is only resonant with the $x$ band. This can also be observed in the polarizability tensor plot on the bottom right where all matrix elements of the polarizability tensor are zero except $\alpha_{xx}$ for almost all $\mathbf q$. Indeed, the structure of $\alpha_{xx}(\mathbf q)$ remains unchanged. This confirms the picture that gapping the bands in the band-structure picture is equivalent to an effective two-level system description in the polarizability tensor picture.

\section{Linear optical elements}
\label{Sec3}
%%%%%%%%%%%%%%%%%%%%%%%%%%%%%%%%%%%%%%%%%%%%
The transmitted field amplitude through a subwavelength array is computed by plugging the result of Eq.~\eqref{eq:beta} into Eq.~\eqref{eq::farfield}. The dipole contribution then reads
\begin{equation}
\label{eq:Edip}
  \mathbf{E}_\text{dip}^{(+)}(Z)  = -\frac{3\pi\Gamma_0}{k_0 }\left[\sum_{j=1}^{\mathcal{N}} \mathbf{G}^\text{far}(Z \hat{e}_z-\mathbf{r}_j) \right]\boldsymbol{\mathcal{M}}^{-1}\mathbf{ E}_{\text{in}}^{(+)},
\end{equation}
thus simply requiring the estimate of the contribution within the square brackets (which we denote in the following by $\bI$). While generally, for arbitrarily chosen lattice geometries, the dipole field can contain many contributions, we will see in the next subsection that, subwavelength arrays impose the cancelations of all other modes except for the fundamental mode at $\mathbf{q}=0$.
%%%%%%%%%%%%%%%%%%%%%%%%%%%%%%%%%%%%%%%%%%%%%%%%%%%%%%%%%%%%%%%%%%%%
\subsection{Derivation of the 2D transmission matrix}
%%%%%%%%%%%%%%%%%%%%%%%%%%%%%%%%%%%%%%%%%%%%
To estimate the sum in the brackets in Eq.~\eqref{eq:Edip} (see Appendix \ref{AppendixC} for full details of the calculation), we make use of the wavevectors $\mathbf{g}$ in the reciprocal lattice $\Lambda^*$ of the array. Any sum over individual sites in the lattice space $\Lambda$ can then be evaluated by an equivalent sum performed in the reciprocal space $\Lambda^*$. Making use of the general form of the Green's tensor (no far field approximation is yet required) the sum becomes
\begin{equation}
   \bI=\frac{\iu}{2k_0^2 \calA}
  \sum_{\mathbf g\in\Lambda^*}
  \frac{k_0^2\mathds{1}-\mathbf v(\mathbf g,Z)\otimes \mathbf v(\mathbf g,Z)}{\sqrt{k_0^2- g^2}}
  \me^{\iu \sqrt{k_0^2-g^2}\abs{Z}},
\end{equation}
where $\mathbf v(\mathbf g,Z)$ is a three-dimensional vector with components of $\mathbf g$ in the $x$ and $y$ directions and component $\text{sgn}(Z)\sqrt{k_0^2-g^2}$ in the $z$ direction. The main observation comes here by a quick inspection of the exponent $\me^{\iu \sqrt{k_0^2-g^2}\abs{Z}}$. For any subwavelength lattice, only the value $g=0$ can lead to a purely imaginary exponent thus describing a propagating plane wave; otherwise, only evanescent wave solutions are obtained as $g^2>k_0^2$. Therefore, we can simply evaluate the contribution of $\mathbf g=(0,0)$ and obtain
\begin{equation}
   \bI=\frac{\iu}{2k_0 \calA} \begin{pmatrix}
1 & 0 & 0 \\
0 & 1 & 0 \\
0 & 0 & 0
\end{pmatrix}
  \me^{\iu k_0\abs{Z}},
\end{equation}
resulting in a total field (both in reflection, for $Z<0$, and in transmission, for $Z>0$) obtained as a sum between incident and scattered field
\begin{align}\label{eq:electric_field_scattering}
\mathbf{E}^{(+)}(Z)=\left[\mathds{1} \me^{\mi k_\ell Z}+\bS\me^{\mi k_\ell |Z|}\right]\mathbf{E}\ts{in}^{(+)},
\end{align}
where we used $k_\ell\approx k_0$ which is also assumed in the rest of the text. The scattering matrix is defined as (see Appendix \ref{AppendixD} for more details such as generalized expression for oblique incidence)
\begin{equation}
\bS=\mi\frac{\tilde{\Gamma}(0)}{2}[\bM^{-1}]\ts{red}=\mi\frac{\tilde{\Gamma}(0)}{2d^2}\boldsymbol{\alpha}\ts{red},
\end{equation}
being easily connected to the polarizability tensor. Here, both $[\bM^{-1}]_\text{red}$ and $\boldsymbol{\alpha}\ts{red}$ are the reduced $2 \times 2$ matrices obtained by a projection into the $xy$ plane of the array.  This stems from the action of the matrix $\bI$ which effectively reduces the problem to 2 dimensions. Therefore, we can now define the  two-dimensional transmission matrix as
\begin{equation}
   \bT=\mathds{1}+\bS=\mathds{1}+\mi \frac{\tilde{\Gamma}(0)}{2d^2}\boldsymbol{\alpha}\ts{red}.
\end{equation}
Note that the subwavelength nature of the array guarantees that, for normal incidence, even if the dipolar response of the surface can have $z$-polarized components, the emitter far field will only contain the $x$- and $y$-polarization components. This allows an input-output formalism where the action of the array is described by a $2\times 2$ Jones matrix.

\subsection{Polarizers}
\label{Sec3B}
%%%%%%%%%%%%%%%%%%%
%%%%%%%%%%%%%%%%%%%
The two simplest examples of Jones matrices \cite{hecht2017optics}
\begin{equation}
\bT_x = \begin{pmatrix}
    1 & 0 \\
    0 & 0
  \end{pmatrix},\quad \bT_y=\begin{pmatrix}
    0 & 0 \\
    0 & 1
  \end{pmatrix},
\end{equation}
\begin{figure}[b]
  \centering
  \includegraphics[width=0.75\columnwidth]{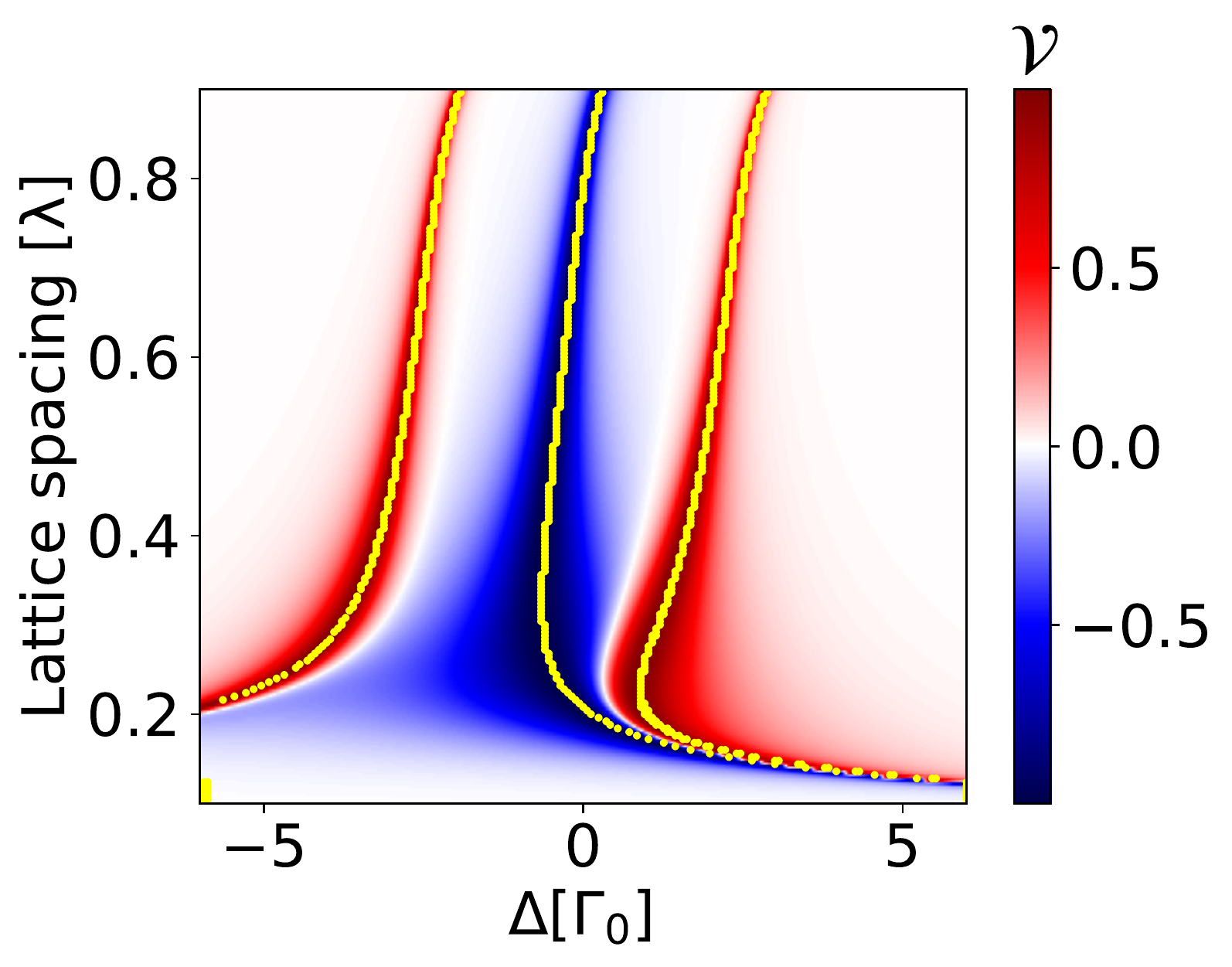}
  \caption{Visibility $\mathcal{V}$ of the $x$- and $y$-polarization components as a function of the laser detuning and lattice spacing (for a square lattice). An external magnetic field in the $x$ direction with magnitude $\mu B_x=3\Gamma_0$ is applied. The blue areas indicate a $y$-polarizer and the red regions an $x$-polarizer implementation. Optimal parameters (corresponding to the resonances of the bands) are highlighted by yellow lines. The condition for obtaining a perfect polarizer, for the $a=0.8\lambda$ case, is when the laser is resonant to the collective surface excitation at the $\Gamma$ point (see also Fig.~\ref{fig:surface_modes}(b)).}
  \label{fig:polarizer}
\end{figure}
describe a polarizer in one direction (either $x$ or $y$) while completely canceling the propagation of the other polarization component. In order to produce such an effect, we make use of a magnetic field oriented along a conveniently chosen direction. The effect of the magnetic field is then to shift the position of the maximum reflectivity points in such a way that the array is not simultaneously reflective/transmissive for both polarization directions. A useful, alternative picture is that of `stripping' the multilevel system of additional levels, until a single two-level system with an appropriately chosen dipole moment orientation is obtained. More precisely, the magnetic field can be used to pull all adjacent undesired levels far from resonance with the incident laser frequency. To see this, we consider the transmission matrix assuming an applied magnetic field $\mathbf B=B_x \hat e_x$ and take the magnetic field towards large values compared to the decay rates such that (see Appendix \ref{AppendixD} for full expression)
\begin{align}
  \label{eq:scattering_matrix_nb}
  \lim_{B_x\to\infty} \bT =\mathds{1}+\frac{\mi \tilde\Gamma(0)/2}{\tilde\Omega^{xx}(0)-\Delta-\iu\tilde\Gamma(0)/2}
  \begin{pmatrix}
    1 & 0 \\
    0 & 0 \\
  \end{pmatrix},
\end{align}
where we used that $\tilde\Omega^{xx}(0)=\tilde\Omega^{yy}(0)$ as it stands for a square lattice. For resonant illumination of the array $\omega_\ell=\omega_0+\tilde\Omega^{xx}(0)$, this describes a $y$-polarizing action in transmission
\begin{align}
  \lim_{B_x\to\infty} \bT =
  \begin{pmatrix}
    0 & 0 \\
    0 & 1 \\
  \end{pmatrix}=\bT_y,
\end{align}
i.e.,~only the $y$ component is transmitted while the $x$ component is perfectly reflected (also compare Figs.~\ref{fig1}(b),(c)). Close to perfect polarizers can be reached as long as the magnetically induced frequency shifts are much larger than $\tilde \Gamma(0)$. Analogously, an $x$ polarizer described by the Jones matrix $\bT_x$ can be implemented by a magnetic field $\mathbf B\parallel{\hat e}_y$.
The action of a polarizer for different lattice constants at a magnetic field strength $\mu B_x=3\Gamma_0$ is illustrated in Fig.~\ref{fig:polarizer}, by plotting the visibility $\mathcal{V}=\left(I_x-I_y\right)/\left(I_x+I_y\right)$ of the intensity components. The particular value of $\mu B_x$ is irrelevant as long as the splitting of the resulting bands is larger than the linewidth of all bands. What can be seen in this plot are the frequencies at which a polarizer in $x$-direction and in $y$-direction performs optimally, indicated by yellow dots. What is also available is information on the linewidth of this polarizing action which is determined by the width of the Lorentzians which emerge when taking horizontal slices of this two-dimensional plot. It is well known that at $0.2\lambda$ and at $0.8\lambda$ there exist resonances at $\Delta=0$ \cite{bettles2016enhanced, Shahmoon2017} which are indeed indicated by yellow dots in the blue-shaded area of the plot. Furthermore, aside from the resonance of the $x$ band indicated by the blue region, one also has the resonances of the hybridized $yz$ bands at positive and negative detunings. Since the $z$ component does not participate for normal incidence, these bands also show the perfect reflection condition for resonant illumination, therefore implementing an $x$ polarizer.\\
\indent Finally, we stress that one is not only restricted to the $x$ or $y$ axes: a general polarizer for an arbitrary axis (e.g.~for $\theta=\pm 45^\circ$) can always be obtained by applying a magnetic field perpendicular to that respective axis. Further more, polarizers for circularly polarized light can be implemented by the application of a magnetic field in $z$ direction. This can be easily understood in the circular basis: By tuning the laser to the distinct resonance of the $\ket +$/$\ket -$ band, a polarization filter for left($-$)/right($+$) circularly polarized light is obtained, respectively (the other component is again perfectly reflected). This is described by the following Jones matrices \cite{hecht2017optics}

\begin{equation}
  \bT_-=\frac{1}{2}
  \begin{pmatrix}
    1 & -\iu \\
    \iu & 1
  \end{pmatrix},\quad
    \bT_+=\frac{1}{2}
  \begin{pmatrix}
    1 & \iu \\
    -\iu & 1
  \end{pmatrix}.
\end{equation}

While these expressions for the transmission matrices are obtained in the Cartesian polarization basis $\{\hat e_x,\hat e_y,\hat e_z\}$, one can equivalently describe the problem in the circular polarization basis $\{\hat e_+,\hat e_-,\hat e_z\}$ with $\hat e_{\pm}=(\hat e_x\pm\iu\hat e_y)/\sqrt{2}$. In this basis, the matrices $\bT_{+/-}$ are analogous to the matrices $\bT_{x/y}$ in the Cartesian basis.

All together, this enables the implementation of highly tunable polarizers for both linearly and circularly polarized light on subwavelength quantum emitter arrays.

%%%%%%%%%%%%%%%%%%%
%%%%%%%%%%%%%%%%%%%
%%%%%%%%%%%%%%%%%%%
%%%%%%%%%%%%%%%%%%%
\subsection{Phase retarders}
\label{Sec3C}
%%%%%%%%%%%%%%%%%%%
%%%%%%%%%%%%%%%%%%%
A generic waveplate (or phase retarder), without losses in either components, can be described by the Jones matrix
\begin{align}
\bT_\phi=
  \begin{pmatrix}
    \me^{\mi\phi_x} & 0 \\
    0 & \me^{\mi\phi_y} \\
  \end{pmatrix},
\end{align}
which imprints a relative phase difference of $\Delta\phi=\phi_x-\phi_y$ onto the outgoing polarization components. Quarter- and half-waveplates would correspond to phase shifts of $\Delta\phi=\pi/2$ and $\pi$, respectively.

\begin{figure}[t]
  \centering
  \includegraphics[width=\columnwidth]{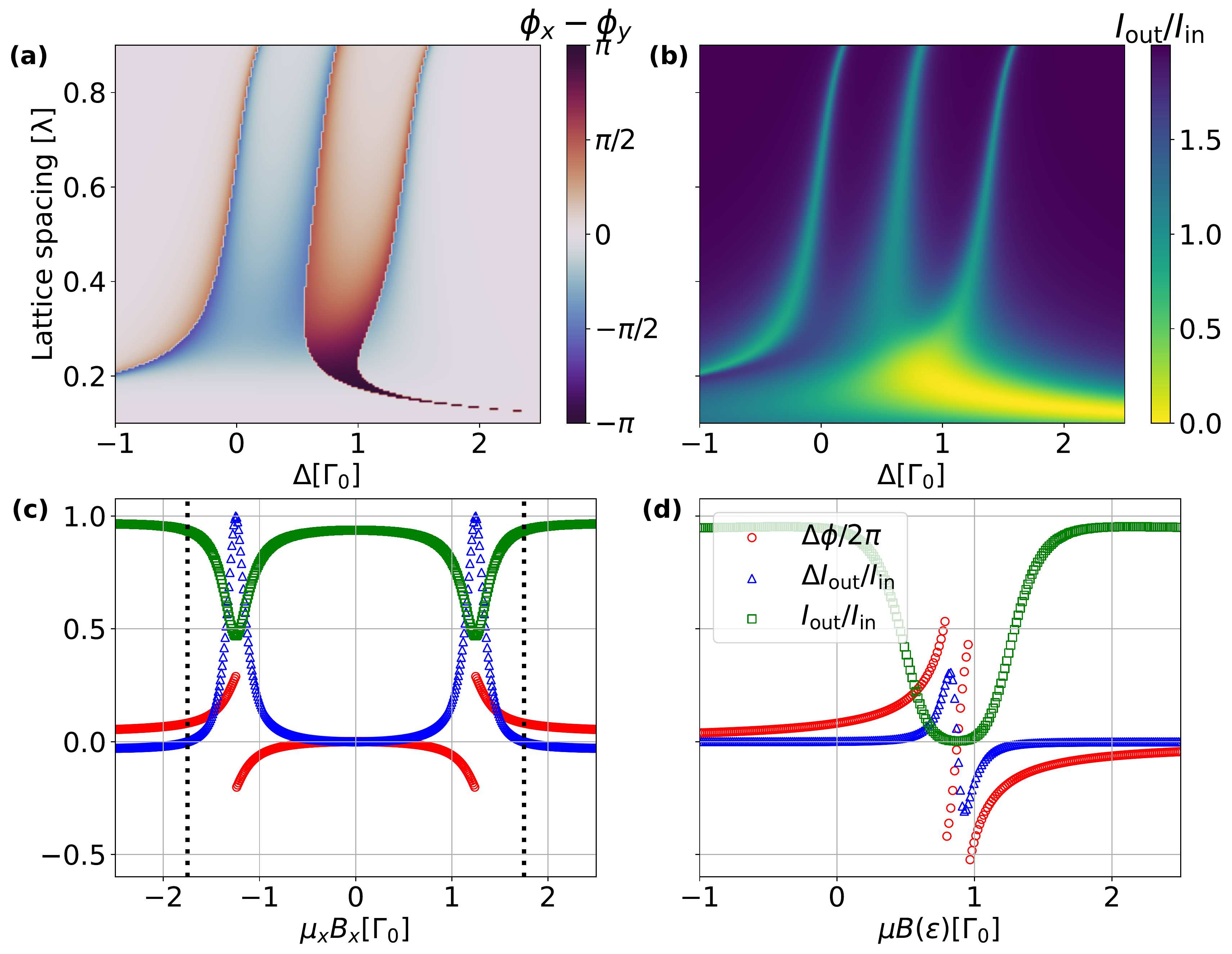}
  \caption{(a) Phase difference $\Delta\phi=\phi_x-\phi_y$ between polarization components of the transmitted light and (b) the corresponding amount of total light transmitted by the emitter array as a function of the lattice spacing and detuning (for a square lattice). The incoming laser beam is equally polarized in $x$ and $y$ direction with no phase difference between them, i.e.,~$\Delta\phi=0$ in the absence of the array. In (c) and (d) we show the phase difference $\Delta\phi$, the total intensity and the intensity difference $\Delta I_\text{out}=I_{x,\text{out}}-I_{y,\text{out}}$ for the $x$- and $y$-polarization components for different external magnetic fields for a square lattice with lattice spacing $a=0.6\lambda$ and detuned by $\Delta=\Gamma_0$ from the bare emitter frequency. The incoming polarization is $(E_{\text{in}, x}, E_{\text{in}, y})=(1,1)$. In (c), the external magnetic field only has an $x$ component for which we show the different quantities. Then in (d), we select the point on the left with $\mu B_x=-1.75\Gamma_0$ (as indicated by the dotted line in (c)), where the intensity difference is minimal and modify both $B_x$ and $B_y$ in a symmetric fashion on the right, i.e., the values on the $x$ axis parameterize $\mu\mathbf B (\varepsilon)/\Gamma_0=(\varepsilon-1.75,\varepsilon,0)$.}
  \label{fig:waveplate}
\end{figure}
Let us now investigate the question of the possible implementation of general waveplates with subwavelength arrays. As analytical considerations are hindered by the cumbersome expressions for the scattering matrix of the array in the presence of external magnetic fields (see Appendix \ref{AppendixD} for detailed expressions) we take recourse to numerical simulations and considerations. The phase retardation effect accompanied by the inevitable loss in transmitted intensity is illustrated in Fig.~\ref{fig:waveplate}, for a square lattice. In Fig.~\ref{fig:waveplate}(a), one can see that a reasonable phase rotation, which is indicated by red or blue color respectively is accompanied by a corresponding reduction in the outgoing intensity with respect to the ingoing intensity on the right indicated by blue. This fundamental problem can be traced back to the fact that the resonances of the array have a Lorentzian structure meaning that the real and imaginary part of the response are strongly dependent on each other. Taking the single-band limit of the expression in Eq.~\eqref{eq:electric_field_scattering}, i.e.,  considering only a single scattering band with effective decay rate $\tilde\Gamma (0)$, frequency shift $\tilde\Omega (0)$ and scattering amplitude $\mathcal{S}$, the phase shift of the transmitted light due to the interaction with the array can be written as
\begin{equation}
  \label{eq:lorentzian}
    \Delta\phi=\arg\left(1+\mathcal{S}\right)=\arctan(\frac{\tilde\Gamma (0)/2}{\tilde\Omega (0)-\Delta}).
\end{equation}
The direction of the phase shift depends on whether the detuning of the laser with respect to the collective resonance is negative or positive. This implies that the contribution of a single band to the phase shift of a polarization component can at most account for $\pi/2$ close to the resonance of the Lorentzian. Around the resonance however, the array is also highly reflective meaning that a large phase shift is accompanied with significant intensity loss. If the laser is resonant with two or more bands, this argument does no longer hold exactly, but the general idea that a larger phase shift also implies more losses due to reflection holds. In this sense, high phase shift differences are always obtained only for very reflecting surfaces. For instance, one can show that for $\Delta\phi=\pi/4$ half of the ingoing intensity of the affected polarization component is always reflected.
\indent We can instead optimize the phase shift difference for an imperfect, lossy but balanced waveplate, where the intensity difference of the incoming polarizations is kept as constant as possible. This would imply that there is simply a prefactor in front of the scattering matrix corresponding to the absolute loss of light intensity due to reflection. An attempt at such an implementation is shown in Fig.~\ref{fig:waveplate}(c) and Fig.~\ref{fig:waveplate}(d), where we select the appropriate magnetic field configuration for a particular lattice and laser frequency in order to define a phase-tunable waveplate. We see analytically, that the $x$ and $y$ component of the magnetic field must be modified in a symmetric fashion in order to keep the intensity difference as constant as possible. Hence, we choose a point where $\Delta I_\text{out}$ is minimal for a magnetic field only pointing in $x$-direction and then parameterize the magnetic field symmetrically around this point $\mu\mathbf  B(\varepsilon)=(-1.75\Gamma_0,0,0)+\varepsilon\Gamma_0\left(\hat e_x+\hat e_y\right)$ (with $\varepsilon$ as parameter). As seen in Fig.~\ref{fig:waveplate}(d) this leads to an extended region where the outgoing intensity is modified significantly, but the intensity difference $\Delta I_\text{out}$ is kept minimal. Thus there exist regions to the left and to the right of the resonance  which provide a magnetic-field tunable waveplate which reduces the intensity in both polarization components in a balanced manner. %Presumably, better regions might exist where the consecutive manipulation of all three components of an externally applied magnetic field are independently varied

%%%%%%%%%%%%%%%%%%%
%%%%%%%%%%%%%%%%%%%
\section{Discussions and extensions}
\label{Sec4}
%%%%%%%%%%%%%%%%%%%
%%%%%%%%%%%%%%%%%%%
A variety of additional factors can perturb the functionality of linear optical elements implemented on two-dimensional subwavelength quantum emitter arrays. We partially address a few of these aspects such as thermal effects and vacancies (which are a limiting factor for atoms trapped in optical lattices \cite{bettles2016enhanced, rui2020asubradiant}) as well as nonlinear effects associated with high intensity driving. We discuss the influence of these effects onto the band shapes and the transmittivity/reflectivity of the array. We also discuss how non-Bravais lattices can be described which we illustrate for a honeycomb lattice.
%%%%%%%%%%%%%%%%%%%%%%%%%%%
%%%%%%%%%%%%%%%%%%%%%%%%%%%
\subsection{Thermal effects}
%%%%%%%%%%%%%%%%%%%%%%%%%%%
%%%%%%%%%%%%%%%%%%%%%%%%%%%

There are different strategies to quantify the action of thermal disturbances in the equilibrium positions of the emitters in the array. In the limit in which the motion is fast compared to the light-matter interaction time scale given by $\Gamma_0^{-1}$, it can be accounted for by averaging over static disorder configurations around the equilibrium positions.
\begin{figure}
  \centering
  \includegraphics[width=\columnwidth]{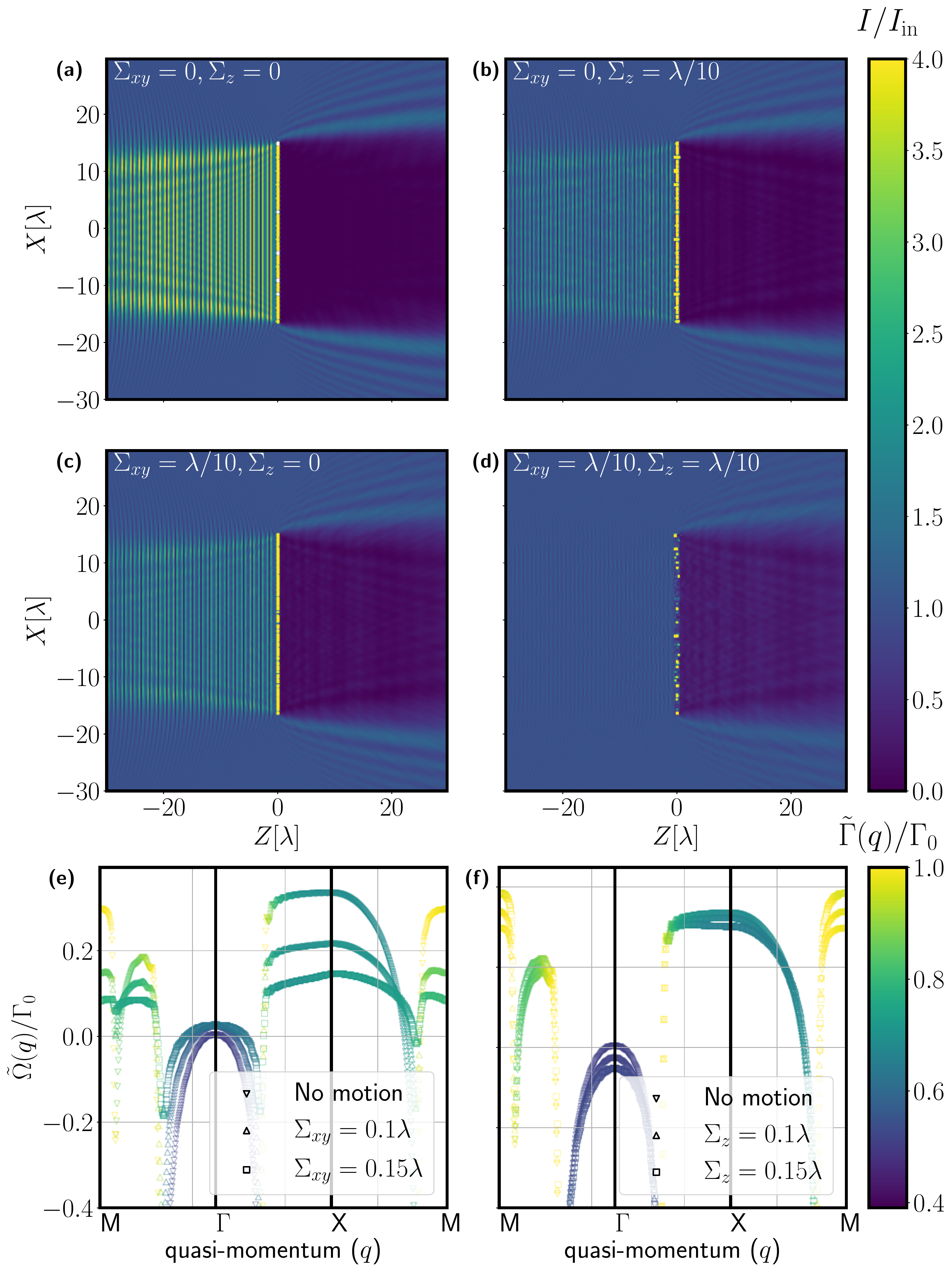}
  \caption{Real space simulations of the electric field intensity of the $x$-polarization component of an incoming field with $(E_{\text{in}, x}, E_{\text{in}, y})=(1,1)$ for a square lattice with $a=0.8\lambda$ and \wfWN. The magnetic field is $\mu B_x=\Gamma_0$. The Gaussian widths of the static disorder are marked by the in-plane $\Sigma_{xy}$ and out-of-plane $\Sigma_z$ symbols. The numerical simulations are performed for $N_c=100$ configurations and then averaged. (e), (f) Effective band diagram calculations including effects of motion. (e) Band diagrams for in-plane motion and (f) band diagram for out-of-plane motion for disorder widths of  $\Sigma=0, 0.1\lambda,   0.15\lambda$, respectively.}
  \label{fig:motion}
\end{figure}
Real space simulations of the effect of motion are shown in Fig.~\ref{fig:motion}. The first observation is that no additional plane wave components aside from the ones propagating in $z$ direction are generated. This is due to the fact that in the infinite-array case the averaging over configurations restores the translational symmetry such that there is no coupling between different momenta if one were to write down an effective description including motion. It is also clear that motion in-plane and out-of-plane have very different effects on the scattered light.\\
\indent Including motion in $z$ direction with Gaussian width $\Sigma_z$ derogates the coherence of the backscattered light. The only effect of coupling to the $z$ component at the $\Gamma$ point in Fig.~\ref{fig:motion}(f) is a small reduction of the frequency since the $xy$ components now hybridize with the $z$ component whose resonance is at a lower frequency at the $\Gamma$ point. Including motion in the array plane with Gaussian width $\Sigma_{xy}$ reduces the polarizer effect more than out-of-plane motion. From Fig.~\ref{fig:motion}(e) we see that including motion in the array plane leads to a broadening of the array response indicated by a larger decay rate for larger $\Sigma_{xy}$. The broadening can be accounted for by considering that motion in the $xy$ plane leads to coupling to other momentum modes at different frequencies. This broadening leads to a reduction in reflectivity. Note that, in order to calculate the modifications of the band diagram we define finite-lattice momentum states $\ket{\mathbf q}=\sum_{\mathbf r_i\in\Lambda}\me^{-\iu\mathbf r_i\mathbf q}\ket{i}$. The band structure is then defined by calculating $E^{c}_{\mathbf q}=\expval{\mathbf \Omega-\iu\mathbf\Gamma/2}{\mathbf q}$ for each configuration $c$ with component matrices  $\mathbf\Omega_{ij}=\mathbf\Omega(\br_{ij})$ and $\mathbf\Gamma_{ij}=\mathbf\Gamma(\br_{ij})$. The averaged band diagram is then taken to be $\bar E_{\mathbf q}=N_c/(\sum_{c}1/E_{\mathbf q}^c)$ which corresponds to an average over the polarizability of the $N_c$ configurations.
%%%%%%%%%%%%%%%%%%%%%%%%%%%%%%
%%%%%%%%%%%%%%%%%%%%%%%%%%%%%%
\subsection{Role of vacancies}
%%%%%%%%%%%%%%%%%%%%%%%%%%%%%%
%%%%%%%%%%%%%%%%%%%%%%%%%%%%%
Vacancies inherently break the translational invariance of the system. They are difficult to treat analytically and numerical studies show that their influence on the radiation pattern is significant.
\begin{figure}
  \centering
  \includegraphics[width=\columnwidth]{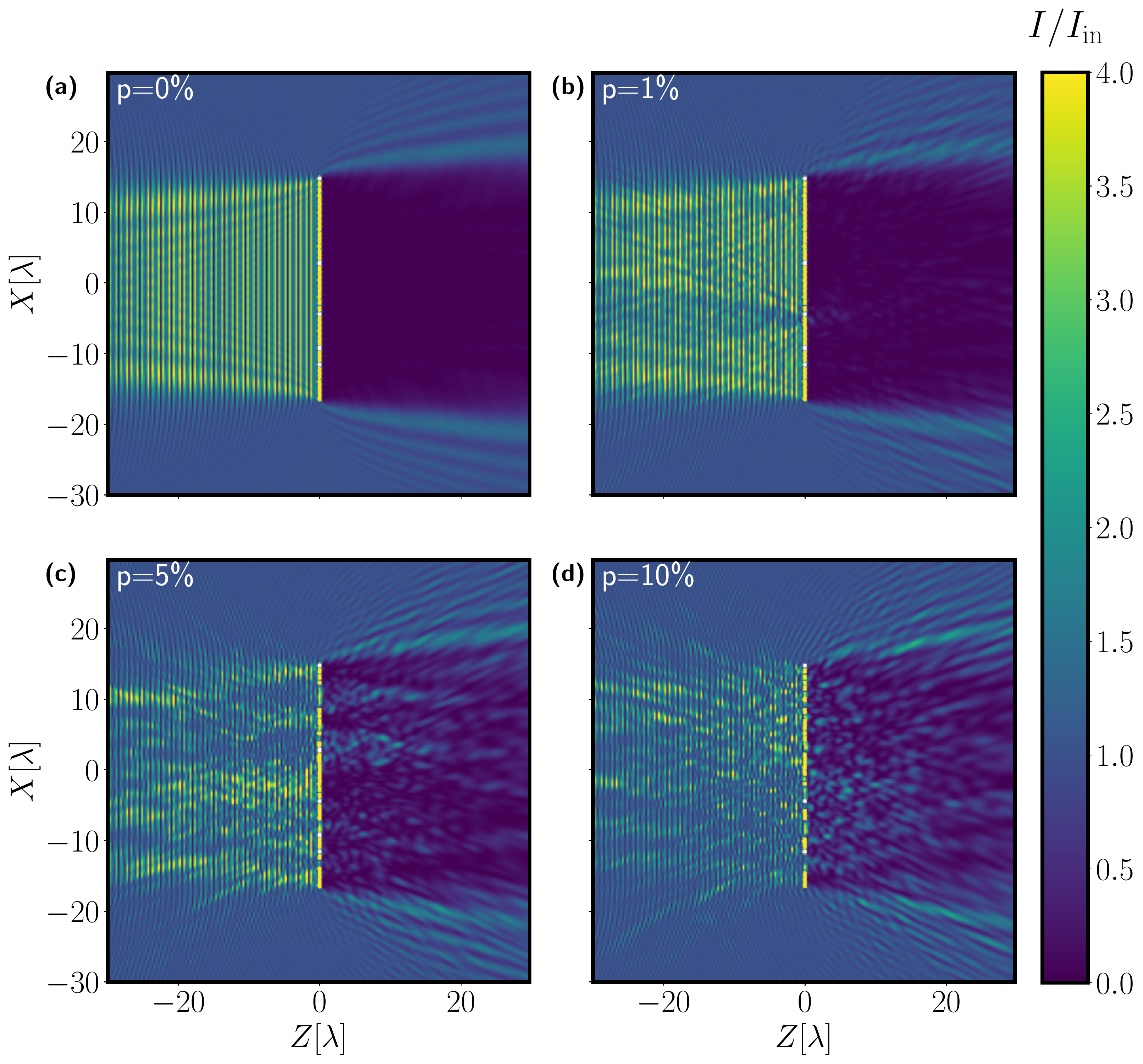}
  \caption{Simulations of the field intensity accounting for single configurations where the lattice has a vacancy randomly located at any lattice site with a probability $p$. (a) $0\%$, (b) $1\%$, (c) $5\%$, and (d) $10\%$. The incoming polarization is $(E_{\text{in}, x}, E_{\text{in}, y})=(1,1)$ and all other parameters are kept as in Fig.~\ref{fig:motion}.}
  \label{fig:vacancies}
\end{figure}
The effect of the breaking of translational symmetry can immediately be seen in Fig.~\ref{fig:vacancies} where we simulate the scattering of light on lattices with an emitter missing with probability $p$ (from $0\%$ to $10\%$) at each lattice site. In particular, for larger vacancy densities a lot of plane-wave components aside from the one propagating in $z$ direction are generated. Since there is usually no averaging over different vacancy configurations, this implies that vacancies deprecate the polarizer effect in a completely different manner from motional disorder which does not break translational invariance. Starting from the translationally invariant picture, the vacancies induce coupling between the previously decoupled momentum modes. This leads to indirect driving of other modes coupled to the laser driven mode which then emit light into free space.

%%%%%%%%%%%%%%%%%%%%%%%%%%%%%%
%%%%%%%%%%%%%%%%%%%%%%%%%%%%%%
\subsection{Beyond the linear regime}
%%%%%%%%%%%%%%%%%%%%%%%%%%%%%%
%%%%%%%%%%%%%%%%%%%%%%%%%%%%%%

\begin{figure}
  \centering
  \includegraphics[width=\columnwidth]{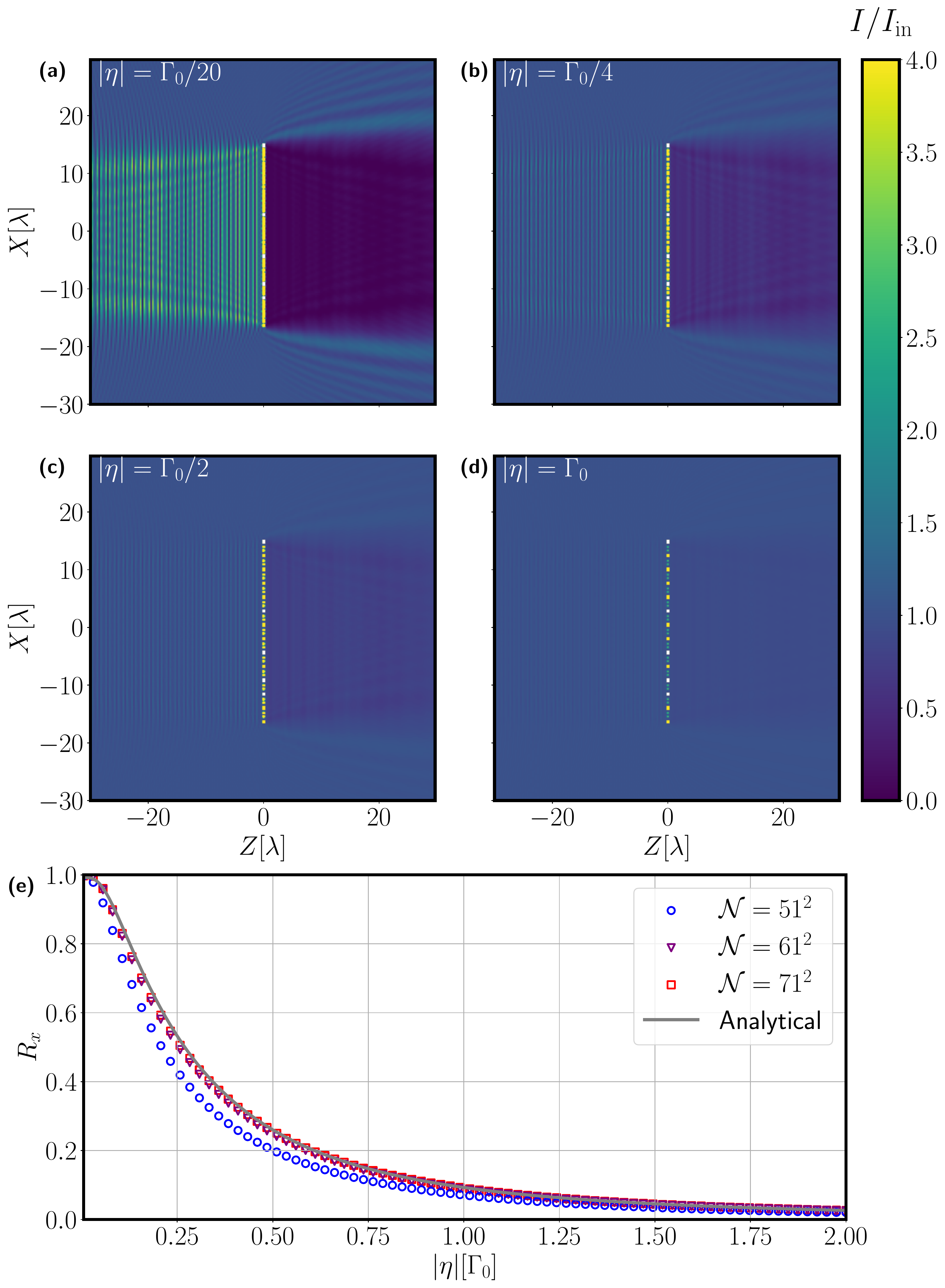}
  \caption{Numerical simulations in the regime reaching beyond linear optical response, for a strong laser drive of the $x$-polarization component in the case where a $y$-polarizer is implemented. Since the drive has the same magnitude everywhere, $\abs{\eta}$ is used as a measure for the nonlinearity. (a) $|\eta|=\Gamma_0/20$, (b) $|\eta|=\Gamma_0/4$,  (c) $|\eta|=\Gamma_0/2$, and (d) $|\eta|=\Gamma_0$. The calculation is performed in real space for a square lattice with $a=0.8\lambda$ also \WN{}. (e) Numerical calculation of the reflectivity of the array $R_x$ for the $x$ polarization for different array sizes ($\mathcal{N}=51^2,61^2,71^2$). It is compared to the analytical expression (obtained from the steady-state solution of Eqs.~\eqref{eqs:nonlin}) as a function of the driving strength $\abs{\eta}$. }
  \label{fig:nonlinear}
\end{figure}
The main focus of the paper has been devoted to the linear response of the array, i.e., the regime in which the emitters are far from saturation. Let us now consider a strong laser drive which can considerably modify the equations of motion of the emitter amplitudes, mainly leading to a power broadening effect responsible with a decrease of the overall reflectivity at higher incoming intensities. Let us first consider the instructing case of a single emitter, driven by an arbitrarily large external light source. Considering only a single polarization degree of freedom, the equations of motion for the coherence $\beta$ and population inversion variable $\beta^z=\expval{\sigma^\dagger\sigma-\sigma\sigma^\dagger}$ are given by
\begin{subequations}
\begin{align}
\dot{\beta}&=-\left(\frac{\Gamma_0}{2}-\mi\Delta\right)\beta+\mi\beta^z\eta,\\
\dot{\beta}^z&=-\Gamma_0 (\beta^z+1) -4\eta\,\text{Im}(\beta).
\end{align}
\end{subequations}
Assuming steady-state, this yields a simple expression of the coherence
\begin{align}
\label{eq:nonlinearbeta}
\beta=\frac{\mi\eta(\Gamma_0/2+\mi\Delta)}{\Gamma_0^2/4+\Delta^2}\beta^z=-\beta^\text{lin}\beta^z,
\end{align}
where we denote by $\beta^\text{lin}$ the response in the linear regime, where $\beta^z$ is close to $-1$ (low excitation regime). The population inversion is given by
\begin{align}
\beta^z=-\frac{1}{1+\frac{2\eta^2}{\Gamma_0^2/4+\Delta^2}},
\end{align}
showing that, with increasing saturation of the emitter, its coherence gets reduced. Assuming moderately high driving intensity, the first correction reads
\begin{align}
\beta\approx \left[1-\frac{2\eta^2}{\Gamma_0^2/4+\Delta^2}\right]\beta^\text{lin},
\end{align}
which is responsible for the Kerr effect where the polarizability of the emitter depends on the intensity of the applied light field. The overall effect on the emitter's response to the external stimulation is then a simple reduction in the overall reflectivity by the reduction in its coherence.\\
\indent Let us now extend the analysis to the full two-dimensional emitter arrays case (also see Refs.~\cite{bettles2020quantum,parmee2021bistable} for related works on nonlinear effects in quantum emitter arrays). A full classical real-space simulation allows to identify the effect of higher driving power as an inhibitor for the overall reflectivity as seen in the progression from Fig.~\ref{fig:nonlinear}(a) to Fig.~\ref{fig:nonlinear}(d) corresponding to an increase of $\eta$ from $\Gamma_0/20$ to $\Gamma_0$. Note that we assume a regime where the applied magnetic field shifts the $y$ resonance very far from the $x$ resonance such that a simplified two-level description fully captures the dynamics of the system. Analytically, we make the same observation as in Sec.~\ref{Sec2C}, that under normal incidence conditions, the emitters respond the same. This allows us to make a mean-field approach and identify $\beta_j=\beta$. The equations of motion become
\begin{subequations}
\label{eqs:nonlin}
\begin{align}
  \dot{\beta}&=-\left(\frac{\Gamma_0}{2}\!-\mi\Delta\right)\!\beta\!
                 +\mi\beta\beta^z\!\left(\tilde{\Omega}(0)\!-\mi\frac{\tilde{\Gamma}(0)\!-\!\Gamma_0}{2}\right)\!+\mi\eta\beta^z,\\
  \dot{\beta}^z&=-\Gamma_0(\beta^z+1)-2(\tilde{\Gamma}(0)-\Gamma_0)|\beta|^2-4\eta\,\text{Im}(\beta).
\end{align}
\end{subequations}
We calculate the results of the equations above in steady-state to obtain an expression for $\beta$ (and thus a renormalized reflectivity) and compare the predicted reflectivity with that of full numerics for real-space scattering. For a relatively moderate array size of $71\times 71$ emitters, the exact numerics converge and agree with the reflectivity derived from the equations above, as illustrated in Fig.~\ref{fig:nonlinear}(e). The main observed effect is the reduction in reflectivity (increase in transmission) with increasing driving power, which can be simply explained by the reduction in the dipole strength predicted by Eq.~\eqref{eq:nonlinearbeta} combined with a modification of the collective interactions due to the nonzero occupancy.\\

%%%%%%%%%%%%%%%%%%%%%%%%%%%%%%
%%%%%%%%%%%%%%%%%%%%%%%%%%%%%%
\subsection{Beyond square lattice: honeycomb lattice}
\begin{figure}[t]
  \centering
  \includegraphics[width=1.0\columnwidth]{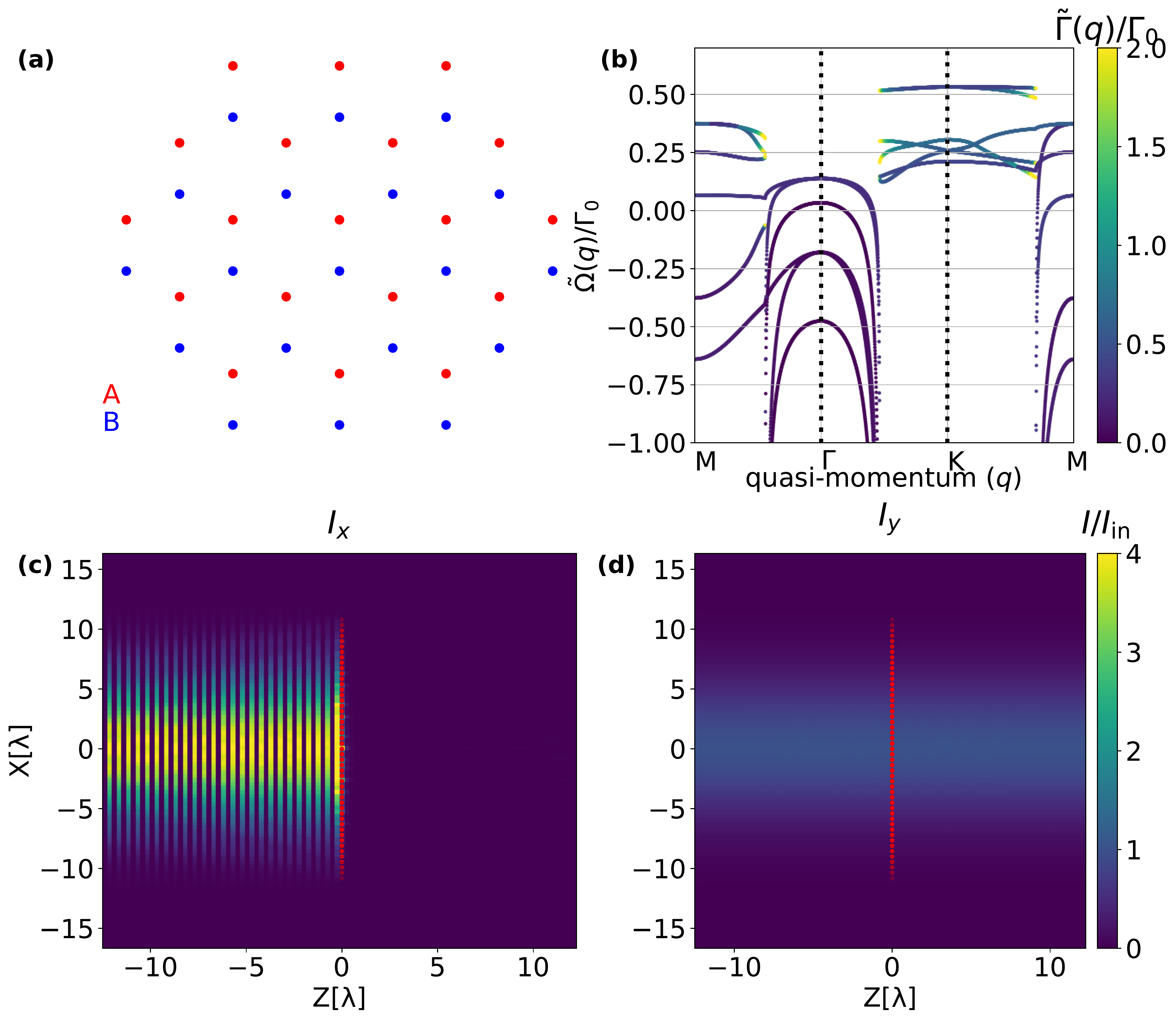}
  \caption{Implementation of a polarizer on a honeycomb lattice with nearest-neighbor spacing $0.9\lambda$. (a) The honeycomb lattice is described by a combination of two triangular (Bravais) sublattices labeled by A and B. (b) Band structure for such a lattice of three level systems (without magnetic field).  (c) and (d)  Intensities of the polarization components. The laser is detuned by $\Delta=-0.18\Gamma_0$ with respect to the natural transition frequency. A magnetic field of strength $\mu B_x=5\Gamma_0$ is applied in $x$-direction.}
  \label{fig:hex}
\end{figure}

%%%%%%%%%%%%%%%%%%%%%%%%%%%%%%
%%%%%%%%%%%%%%%%%%%%%%%%%%%%%%
Let us now exemplify the extension of our formalism and result to more complicated lattice structures, in particular numerically simulating a two-dimensional honeycomb lattice (as illustrated in Fig.~\ref{fig:hex}(a)). The underlying Bravais lattice of the honeycomb lattice is a triangular lattice with a two-atom unit cell. For non-Bravais lattices like the honeycomb lattice, an additional sublattice degree of freedom must be introduced, which we denote by $\nu$. The modification of the theory is detailed in Appendix \ref{AppendixE} and it mainly consists of a recalculation of the Green's tensor now including the additional degree of freedom. For a particular configuration chosen with a nearest-neighbor spacing of $0.9\lambda$, the intensity pattern seen in Fig.~\ref{fig:hex}(c) shows perfect reflection of the $x$ component, while the $y$ component is fully transmitted as seen in Fig.~\ref{fig:hex}(d). The operating conditions were set such that $\Delta=-0.18\Gamma_0$ with respect to the natural transition frequency and a magnetic field of strength $\mu B_x=5\Gamma_0$ is applied.\\
%%%%%%%%%%%%%%%%%%%%%%%%
%%%%%%%%%%%%%%%%%%%%%%%%%%%

\section{Conclusions}

We have investigated the applicability of subwavelength quantum emitter arrays as optical elements in the linear regime. To this end, we have highlighted the necessary steps to first derive the characteristics of the excitation modes on the two-dimensional array and then quantify the linear optical response via the polarizability tensor of the array. We then provided a general analytical expression for the transmission matrix which connects the polarization components of the outgoing, transmitted field to the ones of the incoming field. The subwavelength nature of the array, combined with the particular case of normal incident driving, allowed for a reduction to an exact two-by-two matrix formulation for this scattering problem. The possibility to design the array geometry, in combination with external control through an applied magnetic field allows for the implementation of various Jones matrices such as polarizers and phase retarders. We have also provided some numerical and analytical treatment of various effects that might be detrimental to the efficient operation of such metasurfaces, ranging from thermal motion to disorder and to optical nonlinearities induced by high intensity operation.\\
\indent Such subwavelength arrays might find other applications as, for example, bilayers of emitter arrays  present an interesting platform for cavity quantum electrodynamics with extremely narrow and strongly frequency-dependent linewidths \cite{cernotik2019cavity, reitz2022cooperative}. Another aspect is that of imprinted spin patterns realizable via a spatially varying magnetic field. A possibility is that these spin patterns could be used to implement more generic optical elements such as perfect waveplates. Similar ideas have been put forward for classical metasurfaces~\cite{arbabi2015dielectric}.\\

% None shall float past this point!
\FloatBarrier

\textbf{Acknowledgments} -- We acknowledge financial support from the Max Planck Society and the Deutsche Forschungsgemeinschaft (DFG, German Research Foundation) -- Project-ID 429529648 -- TRR 306 QuCoLiMa
(``Quantum Cooperativity of Light and Matter''). 

\bibliographystyle{apsrev4-1-custom}
\bibliography{references}

\onecolumngrid

\appendix

%%%%%%%%%%%%%%%%%%%%%%%%%%%%%%%%%%%%%%%%%%%%%%%%%%%%%%%
\section{Dyadic Green's function}
\label{AppendixA}

The free-space photonic Green's tensor evaluated at the resonance of the emitters $\omega_0=ck_0$  is given by
\begin{align}
\label{definitiongreen_app}
\mathbf{G}(\mathbf{R})=\left(\mathds{1}+\frac{1}{k_0^2} \nabla\otimes\nabla\right)\frac{\me^{\mi k_0R}}{4\pi R}-\frac{\mathds{1}}{3k_0^2}\delta(\mathbf{R}),
\end{align}
where $\otimes$ denotes the dyadic product, $R=|\mathbf{R}|$ and the last term removes the divergence on the self interaction terms at $\mathbf{R}=0$. This can be expressed more explicitly as
\begin{align}
\mathbf{G}(\mathbf{R})=\frac{\me^{\mi k_0 R}}{4\pi k_0^2}\left[\left(\frac{k_0^2}{R}+\frac{\mi k_0}{R^2}-\frac{1}{R^3}\right)\mathds{1}+\left(-\frac{k_0^2}{R}-\frac{3\mi k_0 }{R^2}+\frac{3}{R^3}\right)\frac{\mathbf{R}\otimes\mathbf{R}}{R^2}\right]-\frac{\mathds{1}}{3k_0^2}\delta(\mathbf{R}),
\end{align}
where $\mathds{1}$ is a $3\times 3$ identity matrix, the vector $\mathbf{R}$ has components $X$, $Y$ and $Z$ and the dyadic product is explicitly given by the following matrix
\begin{align}
\mathbf{R}\otimes\mathbf{R}=\begin{pmatrix}
X^2 & XY &  XZ\\
YX & Y^2 & YZ\\
ZX & ZY & Z^2
\end{pmatrix}.
\end{align}
In the far field, only terms falling off as $1/R$ are contributing and the far field Green's tensor can be expressed in matrix form as
\begin{align}
\mathbf{G}^{\text{far}}(\mathbf{R})= \frac{\me^{\mi k_0 R}}{4\pi R}\begin{pmatrix}
1-X^2/R^2 & -XY &  -XZ\\
-YX &1-Y^2/R^2 & -YZ\\
-ZX & -ZY & 1-Z^2/R^2
\end{pmatrix}.
\end{align}
Notice that for the case where the only component is $Z$, the matrix can be reduced to a $2\times 2$ diagonal matrix with $\me^{\mi k_0 Z}/(4\pi Z)$ on the diagonal. Furthermore, for the scattering problem considered here, a Fourier decomposition of the Green's tensor with respect to the in-plane wave vector components $\bq=(q_x,q_y)$ is useful. Decomposing also $\mathbf R=(\br_\parallel,Z)$ into lattice plane and out-of-plane coordinate, one can make use of the Weyl expansion \cite{novotny2006principles}
\begin{align}
\frac{\me^{\mi k_0 R}}{R}=\frac{\mi}{2\pi}\int \dd \bq\, \frac{1}{q_z}\me^{\mi \bq\cdot \br_\parallel}\me^{\mi q_z |Z|},
\end{align}
with $q_z=\sqrt{k_0^2-q^2}$. The Green's tensor can now be written as
\begin{align}
\label{eq:greenfourier}
\mathbf{G}(\mathbf{R})=\frac{\mi}{8\pi^2 k_0^2}\int \dd\bq \,\frac{k_0^2\mathds{1}-\mathbf v(\mathbf q,Z)\otimes\mathbf v(\mathbf q,Z)}{q_z}\me^{\mi \bq\cdot \br_\parallel}\me^{\mi q_z |Z|},
\end{align}
where we have introduced the function $\mathbf v(\mathbf q,Z)=(q_x,q_y,\text{sgn}(Z)\sqrt{k_0^2-q^2})$. Notice that the expansion above includes the divergent self-interaction (i.e., we have dropped the delta function term).

\section{Master equation formalism}
\label{AppendixB}

The master equation describing the dynamics of the laser-driven quantum emitter array reads
\begin{align}
\frac{\dd \rho}{\dd t}=\mi [\rho, \mathcal{H}]+\sum_{j,j',\alpha,\alpha'}\Gamma^{\alpha, \alpha'}(\mathbf{r}_{jj'})\left[\sigma_{j,\alpha}^{\phantom{\dagger}}\rho\sigma_{j',\alpha'}^\dagger-\frac{1}{2}\left\{\sigma_{j,\alpha}^\dagger\sigma_{j',\alpha'}^{\phantom{\dagger}},\rho\right\}\right],
\end{align}
with the Hamiltonian
\begin{align}
\label{hamarray}
\mathcal{H}=-\sum_{j,\alpha}\Delta \sigma_{j,\alpha}^\dagger\sigma_{j,\alpha}^{\phantom{\dagger}}+\sum_{j,\alpha}\left(\eta_\alpha (\mathbf{r}_j) \sigma_{j,\alpha}^\dagger+\eta_{\alpha}^* (\mathbf{r}_j)\sigma_{j,\alpha}^{\phantom{\dagger}}\right)+\sum_{j, j', \alpha,\alpha'}\Omega^{\alpha,\alpha'}(\mathbf{r}_{jj'})\sigma_{j,\alpha}^\dagger\sigma_{j',\alpha'}^{\phantom{\dagger}}+\mathcal{H}_{\text{B}},
\end{align}
where $\eta_{\alpha} (\mathbf{r}_j)=\eta_\alpha \me^{\mi \bk_\parallel \cdot \br_j}$ is the generalized position-dependent Rabi frequency for oblique incidence with wave vector $\bk_\parallel$ parallel to the array plane. $\mathcal{H}_{\text{B}}$ is the Hamiltonian which describes the effect of a magnetic field. In the Cartesian basis which we mainly use this Hamiltonian reads
\begin{equation}
  \label{eq:magnetic_hamiltonian}
  \mathcal{H}_{\text{B}}=\iu\varepsilon_{\alpha_1,\alpha_2,\alpha_3}\mu_{i,\alpha_1}B_{i,\alpha_1}\ket{i,\alpha_2}\bra{i,\alpha_3},
\end{equation}
with Einstein sum convention where $\alpha_i=x,y,z$ and $\mu_{i,\alpha}$, $B_{i,\alpha}$ are the Cartesian components of the magnetic dipole moments and the magnetic field at emitter $i$, respectively. $\varepsilon_{ijk}$ is the Levi-Civita symbol.

\section{Fourier space expressions for far-field Green's tensor and decay rates}
\label{AppendixC}

Let us now move to the Fourier domain. The dipole-scattered field can be expressed as an integral over all wave vectors in the Brillouin zone $\mathcal{B}$
\begin{equation}\label{eq:A_scattered_field}
  \mathbf{E}^{(+)}_{\text{dip}}({\mathbf R})=\frac{3\pi\Gamma_0}{k_0d}\int\limits_{\mathbf q\in\mathcal{B}}\dd\mathbf q\,\me^{\iu\mathbf q\cdot\mathbf r_\parallel}\widetilde{\mathbf G}(\mathbf q;Z)\boldsymbol\beta_{\mathbf q},
\end{equation}
where $\widetilde{\mathbf G}(\mathbf q;Z)=\sum_{\mathbf r_\parallel\in\Lambda}\me^{-\iu\mathbf q\cdot\mathbf r_\parallel}\mathbf G(\mathbf R)$ is the (discrete) 2D lattice Fourier transform in the first two arguments of the Green's tensor. We can connect the discrete $\widetilde{\mathbf G}(\mathbf q;Z)$ and the continuous Fourier transform $\bar{\mathbf G}(\mathbf q;Z)=\frac{1}{\left(2\pi\right)^2}\int_{\br_\parallel\in\mathbb{R}^2}\diff{\br_\parallel}\me^{-\iu\bq\cdot\br_\parallel}\mathbf G(\mathbf R)$ by making use of the Poisson summation formula in two dimensions $\sum_{\br_\parallel\in\Lambda}f(\br_\parallel)=\frac{\left(2\pi\right)^2}{\mathcal A}\sum_{\mathbf g\in\Lambda^*}\bar f(\mathbf g)$ where $\mathcal A$ is the area of the unit cell. To recast the lattice transform of the Green's tensor into a sum over reciprocal lattice vectors $\mathbf g$, we identify $f(\mathbf r_\parallel)=\me^{-\iu\mathbf q\cdot\mathbf r_\parallel}\mathbf G(\mathbf R)$ in the Poisson sum formula to obtain
\begin{equation}
  \label{eq:Poisson_sum}
 \widetilde{\mathbf G}(\mathbf q;Z)=\frac{(2\pi)^2}{\calA}\sum_{\mathbf g\in\Lambda^*}\bar{\mathbf G}(\mathbf q+\mathbf g;Z).
\end{equation}
The expression for the continuous Fourier transform can be identified from the Weyl expansion Eq.~\eqref{eq:greenfourier}
\begin{equation}
  \label{eq:analytical_gf}
  \bar{\mathbf G}(\mathbf q;Z)=\frac{\iu}{8\pi^2}\left(\mathds{1}-\frac{\mathbf v(\mathbf q,Z)\otimes \mathbf v(\mathbf q,Z)}{k_0^2}\right)\frac{\me^{\iu q_z|Z|}}{q_z}.
\end{equation}
 Instead of a sum over the real-space lattice, $\widetilde{\mathbf G}$ is now represented via a reciprocal lattice summation

\begin{equation}
  \label{eq:lattice_ft}
  \widetilde{\mathbf G}(\mathbf q;Z)=\frac{\iu}{2\calA}
  \sum_{\mathbf g\in\Lambda^*}
  \left(\mathds{1}-\frac{\mathbf v(\mathbf q+\mathbf g,Z)\otimes \mathbf v(\mathbf q+\mathbf g,Z)}{k_0^2}\right)
  \frac{\me^{\iu \sqrt{k_0^2-(\mathbf q+\mathbf g)^2}\abs{Z}}}{\sqrt{k_0^2-(\mathbf q+\mathbf g)^2}}.
\end{equation}

Notably, the geometric prefactor in front of the plane wave ensures that the polarization of the outgoing plane wave is perpendicular to the propagation direction. It is thus a projector onto the plane orthogonal to $\mathbf v(\mathbf q+\mathbf g, Z)$.

The fundamental argument for the directional character of the light scattered from subwavelength arrays is the fact that the exponential term $\me^{\iu \sqrt{k_0^2-(\mathbf q+\mathbf g)^2}\abs{Z}}$ becomes evanescent for $\mathbf g\neq 0$ when $a<\lambda/2$ (this condition depends on the angle of incidence, with $a<\lambda/2$ for full oblique and $a<\lambda$ for normal incidence) so that only the zeroth summand in the reciprocal lattice sum remains

\begin{equation}
  \label{eq:lattice_ft_zero}
  \widetilde{\mathbf G}^{\text{far}}(\mathbf q;Z)\approx\frac{\iu}{2\calA}\frac{\me^{\iu q_z\abs{Z}}}{q_z}
  \left(\mathds{1}-\frac{\mathbf v(\mathbf q,Z)\otimes\mathbf v(\mathbf q,Z)}{k_0^2}\right).
\end{equation}
For perpendicular incidence of the laser only the $\bq=0$ component will be selected and this simply becomes
\begin{align}
\widetilde{\mathbf G}^{\text{far}}(0;Z)=\frac{\mi}{2\calA}\frac{\me^{\mi k_0 \abs{Z}}}{k_0}\begin{pmatrix}
1 & 0 & 0 \\
0 & 1 & 0 \\
0 & 0 & 0
\end{pmatrix}.
\end{align}
The same argument as above can be used to deduce the decay rate for subwavelength arrays since the plane-wave term only contributes to the imaginary part if $\mathbf g=0$ so that
\begin{equation}
  \label{eq:decay_analytical}
  \widetilde{\boldsymbol\Gamma}(\mathbf q)\approx\frac{3\Gamma_0}{4\pi}\left(\frac{\lambda^2}{\calA}\right)\frac{k_0}{q_z}\left(\mathds{1}-\frac{\mathbf v^+(\mathbf q)\otimes \mathbf v^+(\mathbf q)}{k_0^2}\right),
\end{equation}
with $\mathbf v^+(\bq)=\mathbf v(\bq,0^+)=(q_x,q_y,\sqrt{k_0^2-q^2})$. This can be used below to simplify the expression for the scattering matrix of the 2D array. Notice that the expression for the discrete Green's tensor does not include the delta function term. Therefore, it suffices for the analytical computation of the far field Green's tensor and of the decay rate matrix but it cannot be used to compute the dipole-dipole collective shifts. The reason for that is that the summation does not converge owing to the divergent self interaction. This can instead be dealt with more easily in real space by canceling any contributions from $\mathbf r_\parallel=0$ or alternatively, via a fluctuation averaging procedure as in Ref.~\cite{Perczel2017}.

\section{Explicit expression of scattering matrix}
\label{AppendixD}

  The matrix $\bM(\mathbf q)=-\Delta\mathds{1}+\tilde{\bO}(\bq)-\mi\tilde{\bG}(\bq)/2$, which appears in Sec.~\ref{Sec2} when solving the linear system for the polarizations of the dipoles on the lattice is explicitly given by (without applied magnetic field)
\begin{equation}
  \bM(\mathbf q)=-\Delta \mathds{1}+\begin{pmatrix}
    \tilde\Omega^{xx}(\mathbf q)-\mi\tilde\Gamma^{xx}(\mathbf q)/2 & \tilde\Omega^{xy}(\mathbf q)-\mi\tilde\Gamma^{xy}(\mathbf q)/2 & 0 \\
   \tilde\Omega^{yx}(\mathbf q)-\mi \tilde\Gamma^{yx}(\mathbf q)/2 & \tilde\Omega^{yy}(\mathbf q)-\mi\tilde\Gamma^{yy}(\mathbf q)/2 & 0\\
    0 & 0 & \tilde\Omega^{zz}(\mathbf q)-\mi\tilde\Gamma^{zz}(\mathbf q)/2
  \end{pmatrix}.
\end{equation}
This matrix is is block-diagonal since there is no coupling to the $z$ component for primitive lattice vectors in the $xy$ plane. Defining the projector into the plane orthogonal to $\mathbf v$ as $\boldsymbol{\mathcal{P}}_{\mathbf v}$, the decay rate matrix for a subwavelength array in Fourier space may be approximated from Eq.~\eqref{eq:decay_analytical}  as $\tilde{\mathbf \Gamma} (\mathbf q)\approx \tilde\Gamma(\mathbf q)\boldsymbol{\mathcal{P}}_{\mathbf v^+ (\mathbf q)}$. A magnetic field $\mathbf B=(B_x,B_y,B_z)$ introduces couplings among the Cartesian components with magnetic dipole moments $\boldsymbol \mu =(\mu_x, \mu_y, \mu_z)$ which adds to the above matrix $  \bM(\mathbf q)$ as
\begin{equation}
 \mathbf{M}_\text{B}=\begin{pmatrix}
    0& -\iu B_z\mu_z  & \iu\mu_yB_y \\
   \iu B_z\mu_z & 0 & -\iu\mu_xB_x\\
     -\iu\mu_yB_y & \iu\mu_xB_x  & 0
  \end{pmatrix}.
 \end{equation}
Including the effect of the magnetic field, the $\mathbf q$ dependent scattering matrix for a subwavelength array may finally be expressed as
\begin{equation}
  \label{eq:scattering_matrix}
  \bS(\mathbf q, Z)=
  \mi \frac{\tilde{\Gamma}(\mathbf q)}{2}
  \boldsymbol{\mathcal{P}}_{\mathbf v(\mathbf q,Z)}\left[\begin{pmatrix}
    -\Delta+\tilde\Omega^{xx}(\mathbf q) & \tilde\Omega^{xy}(\mathbf q)-\iu B_z\mu_z & \iu\mu_yB_y \\
    \tilde\Omega^{yx}(\mathbf q)+\iu B_z\mu_z & -\Delta+\tilde\Omega^{yy}(\mathbf q) & -\iu\mu_xB_x\\
    -\iu\mu_yB_y & \iu\mu_xB_x & -\Delta+\tilde\Omega^{zz}(\mathbf q)
  \end{pmatrix}
  -\iu  \frac{\tilde{\Gamma}(\mathbf q)}{2}  \boldsymbol{\mathcal{P}}_{\mathbf v^+ (\mathbf q)}
\right]^{-1}\boldsymbol{\mathcal{P}}_{\mathbf v^+(\mathbf q)}.
\end{equation}
This still depends on the sign of $Z$ as for non-normal incidence, the reflected and transmitted waves have different propagation directions.

Generally, the matrix inversion in the equation above leads to a cumbersome expression for the scattering matrix. A simplified expression can be obtained by considering a square lattice at normal incidence $\mathbf q = 0$ which corresponds to the situation mostly considered in the main text of the manuscript. Additionally  only considering a magnetic field along the  $x$ direction, $\mathbf B = B_x{\hat e}_x$, the scattering matrix $\bS (0)\equiv \bS$ becomes
\begin{align}
\bS = C\begin{pmatrix}
   - \mu_x^2 B_x^2+\left(\mi\tilde\Gamma (0)/2+\Delta-\tilde\Omega^{xx} (0)\right)\left(\Delta-\tilde{\Omega}^{zz}(0)\right) & \tilde{\Omega}^{xy}(0)\left(\Delta-\tilde{\Omega}^{zz}(0)\right) \\
    \tilde{\Omega}^{xy}(0)\left(\Delta-\tilde{\Omega}^{zz}(0)\right) & \left(\mi\tilde\Gamma (0)/2+\Delta-\tilde\Omega^{xx} (0)\right)\left(\Delta-\tilde{\Omega}^{zz}(0)\right)\\
  \end{pmatrix},
\end{align}
with
\begin{align}
C=\frac{\mi\tilde \Gamma (0)/2}{\mu_x^2 B_x^2 \left(\mi\tilde\Gamma (0)/2+\Delta-\tilde\Omega^{xx} (0)\right)+\left[\left(\mi\tilde\Gamma (0)/2+\Delta-\tilde\Omega^{xx} (0)\right)^2-\tilde{\Omega}^{xy}(0)^2\right]\left(\tilde{\Omega}^{zz}(0)-\Delta\right)},
\end{align}
where we used that $\tilde\Omega^{xx}(0)=\tilde\Omega^{yy}(0)$ for the square lattice. In the limit of $B_x\to\infty$, this leads to the expression for the transmission matrix $\bT=\mathds{1}+\bS$ in Eq.~\eqref{eq:scattering_matrix_nb}.

\section{Generalization to non-Bravais lattices}
\label{AppendixE}

For more complicated lattices that are non-Bravais, an additional sublattice degree of freedom must be introduced since the lattice Fourier transform is only well defined for a Bravais lattice~\cite{Perczel2017}. Let us denote the sublattice degree of freedom by $\nu$.

Indeed, first of all Eq.~\eqref{eq:A_scattered_field} is actually unchanged, but the Green's tensor now includes the sublattice degree of freedom. For the sublattice blocks, one can then more explicitly denote how the lattice Fourier transforms of the blocks differ
\begin{equation}
  \label{eq:Poisson_sum_NB}
  \widetilde{\mathbf G}_{\nu,\nu'}(\mathbf q;Z)=\sum_{\mathbf r_\parallel\in\Lambda}\me^{-\iu\mathbf q\mathbf r_\parallel}\mathbf G(\mathbf r_\parallel+\mathbf b_{\nu,\nu'}+\hat e_z Z),
\end{equation}
where $\mathbf b_{\nu,\nu'}=\mathbf b_{\nu}-\mathbf b_{\nu'}$ for the sublattice vectors $\mathbf b_{\nu}$ which are also the locations of the emitters in the unit cell. For each sublattice block, the Poisson sum formula can now be leveraged to write

\begin{equation}
  \label{eq:Poisson_formula_NB}
  \widetilde{\mathbf G}_{\nu,\nu'}(\mathbf q;Z)=\frac{(2\pi)^2}{\calA}\sum_{\mathbf g\in\Lambda^*}\me^{\iu(\mathbf q+\mathbf g)\cdot\mathbf b_{\nu,\nu'}}\bar{\mathbf G}(\mathbf q+\mathbf g;Z).
\end{equation}
This has some remarkable consequences. One point is the far-field (only taking the $\mathbf g=0$ contribution)
\begin{equation}
  \label{eq:lattice_ft_NB}
  \widetilde{\mathbf G}_{\nu,\nu'}^{\text{far}}(\mathbf q;Z)\approx\frac{\mi}{2\calA}\frac{\me^{\iu q_z \abs{Z}}}{q_z}\me^{\iu\mathbf q\cdot\mathbf b_{\nu,\nu'}}
  \left(\mathds{1}-\frac{\mathbf v(\mathbf q,Z)\otimes\mathbf v(\mathbf q,Z)}{k_0^2}\right),
\end{equation}
which simply implies that the plane waves emitted by the different sublattices interfere in a way that is determined by the spacing between the sublattices.

Furthermore, the matrix consisting of these sublattice phases is Hermitian such that the argument that only the $\mathbf g=0$ component contributes to the decay rate of the band structure for $a<\lambda/2$ can also be applied for multiple sublattices

\begin{equation}
  \label{eq:decay_analytical_appendix}
  \widetilde{\boldsymbol\Gamma}_{\nu,\nu'}(\mathbf q)\approx\frac{3\Gamma_0}{4\pi}\left(\frac{\lambda^2}{\calA}\right)\frac{k_0}{q_z}\me^{\iu\mathbf q\cdot\mathbf b_{\nu,\nu'}}\left(\mathds{1}-\frac{\mathbf v^+(\mathbf q)\otimes \mathbf v^+(\mathbf q)}{k_0^2}\right).
\end{equation}
It is particularly useful now to consider the special case of two sublattices. We can then choose the sublattice vectors as $\mathbf b_1=0$ and $\mathbf b_2=\mathbf b$ so that the sublattice matrix becomes

\begin{equation}
  \label{eq:sublattice_matrix}
  \begin{pmatrix}
    0 & \me^{\iu\mathbf q\mathbf b}\\
    \me^{-\iu\mathbf q\mathbf b} & 0
  \end{pmatrix}.
\end{equation}
This matrix has eigenvalues $\xi_1=0$ and $\xi_2=2$. Since the $\mathbf g=0$ term also contributes in a major way to the frequency component of the Green's function this implies that the diagonalization leads to a very flat, subradiant band and a more dispersive superradiant (with respect to the case without an additional sublattice) band corresponding respectively to the eigenvalues $0$ and $2$.

\end{document}